\PassOptionsToPackage{hyphens}{url}
\PassOptionsToPackage{sort}{natbib}
\documentclass[sageh,times, 10pt]{sagej}
\usepackage[T1]{fontenc}
\usepackage[utf8]{inputenc}
\usepackage{amsmath,amssymb,booktabs,array,tabularx,graphicx,placeins}
\usepackage{microtype}
\usepackage[colorlinks,bookmarksopen,bookmarksnumbered,citecolor=blue,urlcolor=blue]{hyperref}
\setcitestyle{authoryear,round,semicolon,aysep={,},yysep={,},notesep={: }}
\newcommand{\Prb}{\operatorname{Pr}}
\newcommand{\logit}{\operatorname{logit}}
\hypersetup{bookmarksdepth=2,pdftitle={What If They Took the Shot? A Hierarchical Bayesian Framework for Counterfactual Expected Goals},pdfauthor={},pdfkeywords={hierarchical Bayesian modelling, counterfactual comparison, expected goals, uncertainty quantification, football analytics}}
\graphicspath{{figures/}}
\begin{document}
\title{What If They Took the Shot? A Hierarchical Bayesian Framework for Counterfactual Expected Goals}
\author{Mikayil Mahmudlu\affilnum{1} and Oktay Karaku\c{s}\affilnum{2,3} and Hasan Arkada\c{s}\affilnum{3} and Can Ey\"{u}po\u{g}lu\affilnum{2,4}}
\affiliation{\affilnum{1}Department of Mathematics,
Cardiff University, Abacws, Senghennydd Road,
Cardiff CF24 4AG, UK.\\
\affilnum{2}Department of Computer Science and Informatics,
Cardiff University, Abacws, Senghennydd Road,
Cardiff CF24 4AG, UK.\\
\affilnum{3}Dead Ball Analytics Limited, Barry, CF63 2QQ, UK\\
\affilnum{4}Department of Computer Engineering, Turkish Air Force Academy, National Defence University, Istanbul, Türkiye}
\corrauth{Oktay Karaku\c{s}, Department of Computer Science and Informatics,
Cardiff University, Abacws, Senghennydd Road,
Cardiff CF24 4AG, UK.}
\email{karakuso@cardiff.ac.uk}
\begin{abstract}
How would the probability of scoring change if a different player
took the same shot? Although expected goals models describe the
difficulty of scoring opportunities, comparing players on identical
chances requires accounting for uncertainty in their estimated
conversion probabilities. We address this question through a
hierarchical Bayesian framework that evaluates alternative shooters
under the same observed circumstances. Using 35,700 eligible shots
from four leagues during the 2015/16 season, we compare a contextual
benchmark with a primary player-intercept model and a provisional
extension that allows player differences to vary with shooting
conditions. Paired posterior draws preserve shared uncertainty when
individual probability differences are aggregated across
opportunities. Chronological validation demonstrates improved
predictive performance relative to a constant-rate baseline, but
introducing player effects provides negligible additional gains.
Controlled single-seed experiments show that greater information
per player improves estimation accuracy and narrows uncertainty
intervals, although distinguishing small differences remains
difficult and confident conclusions about near-null effects can
become more frequent. Applications involving Giroud--Vardy and
Berardi--Sansone similarly yield estimated differences that are
small relative to their posterior uncertainty. These findings
highlight an important distinction between predicting the outcome
of a shot and establishing whether one player would be more likely
to convert it. The proposed framework makes this distinction
explicit by evaluating alternative shooters on fixed recorded
opportunities while accounting for their contextual exposure.
The resulting comparisons concern conditional conversion rather
than the causal consequences of replacing a player in a match.
\end{abstract}
\maketitle

\section{Introduction: what if they took the shot?}
\label{sec:introduction}

Expected goals (xG) have become an established means of evaluating
scoring opportunities in football by estimating the probability
of a goal from the circumstances surrounding a shot. Rather than
judging an attempt solely by whether it resulted in a goal, xG
recognises that the outcome of any individual shot is uncertain
and that opportunities differ considerably in difficulty
\citep{brechot2020}. This distinction provides a more informative
basis for interpreting attacking performance, but it also raises
a question that conventional predictions for the observed shooter
cannot answer. If two players were presented with exactly the
same opportunity, how confidently could we determine which
would be more likely to score?

\begin{figure}[tb]
\centering\includegraphics[width=\linewidth]{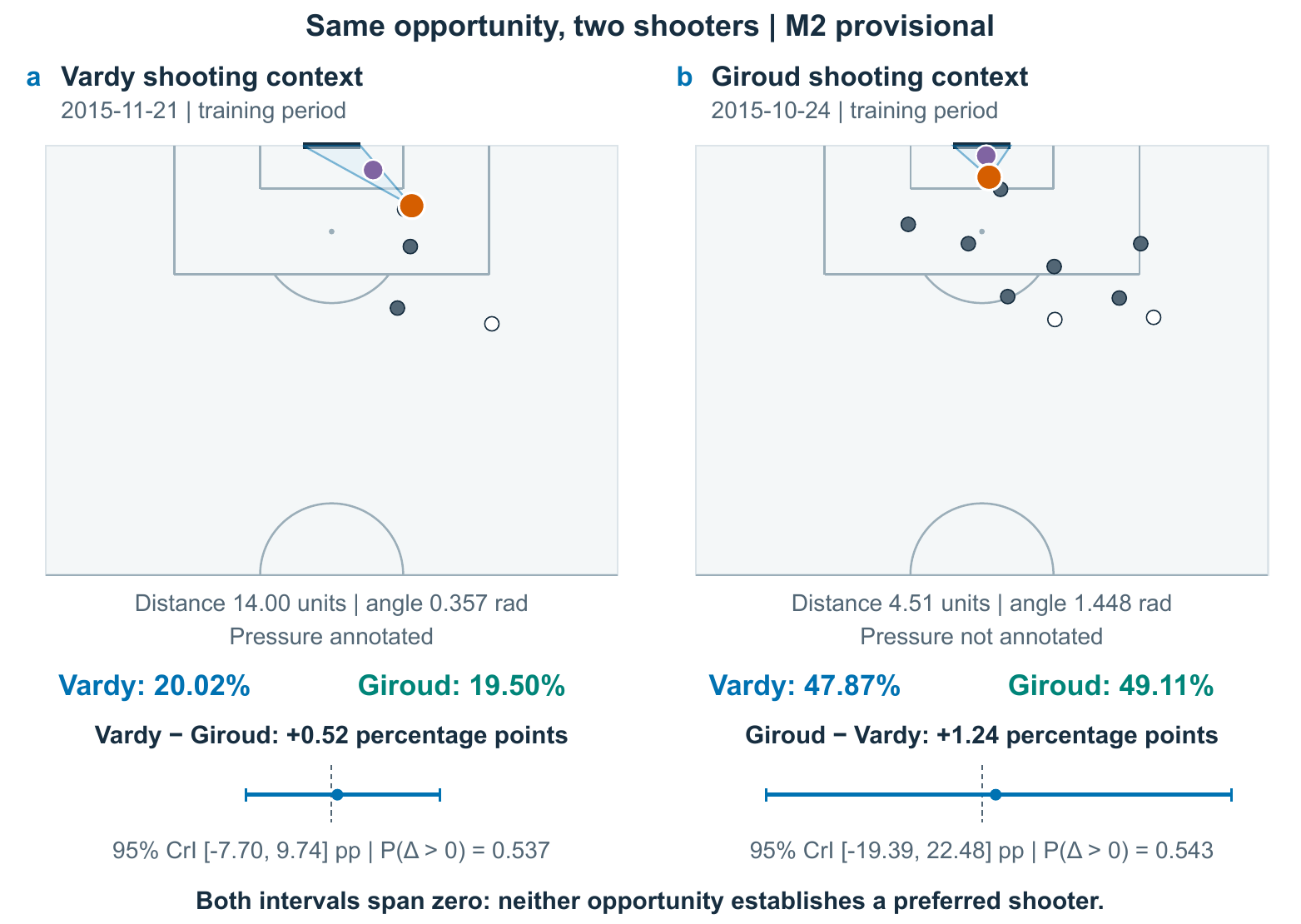}
\caption{
\textbf{Comparing two shooters on the same recorded opportunity.}
The figure illustrates conditional conversion comparisons using
two training-period attempts: (a) Vardy's shot on 21 November
2015 and (b) Giroud's shot on 24 October 2015. Both players
are evaluated under the recorded circumstances using the
provisional context-dependent model M2
(Section~\ref{sec:models}). The half-pitches are rotated
counter-clockwise, with the goal positioned at the top,
while preserving the proportions of the original
$120\times80$ coordinate system. Orange identifies the
shooter, purple the opposing goalkeeper, filled grey
markers other opponents and hollow markers teammates.
The triangle connects the shot location to the goalposts.
The examples were selected retrospectively because they
produced the largest opposing posterior-mean contrasts
among opportunities within both players' training
distance and angle ranges. Their pressure annotations
and body-part/technique combinations also occur in both
players' records, although this marginal overlap does
not establish joint contextual support. Posterior-mean
probabilities, paired 95\% credible intervals and sign
probabilities describe the estimated differences between
shooters. Both intervals include zero, leaving the
preferred shooter unresolved in either example.
The primary player-intercept model produces no opposing
posterior-mean preferences on these opportunity sets.
Source event identifiers document the selected attempts.
}
\label{fig:concept}
\end{figure}

The question is appealing because the opportunities available
to footballers are rarely equivalent. A forward must find or
create space, receive the ball, recognise the possibility of
shooting and execute the attempt, with each stage contributing
to the goals ultimately recorded. Differences in goal totals
therefore reflect more than conversion alone. Comparing Olivier
Giroud with Jamie Vardy on Vardy's recorded shots, for example,
asks how their estimated conversion probabilities differ under
the circumstances Vardy actually encountered. Reversing the
comparison and evaluating both players on Giroud's opportunities
addresses a different question. Neither exercise establishes
whether the alternative player would have reached those positions
or chosen to shoot. Instead, holding the measured circumstances
fixed provides a way to examine differences in estimated
conversion without treating the opportunities themselves as
interchangeable. Figure~\ref{fig:concept} illustrates this idea
using individual attempts.

Existing xG research provides much of the modelling foundation
needed for such comparisons. Shot location, positional information
and other recorded circumstances help characterise opportunity
difficulty, while player-adjusted approaches allow conversion
probabilities to differ between shooters
\citep{anzer_bauer_2021,code_3,code_8,code_2}. Yet the presence
of player effects in a predictive model does not necessarily mean
that the available evidence can distinguish two particular
players. Shooting records vary substantially in size and
composition, goals remain relatively uncommon, and players
encounter different combinations of opportunities. A model may
therefore predict observed outcomes reasonably well while
providing only weak evidence about which of two shooters would
be more likely to convert the same chance. This distinction
between predictive performance and the uncertainty surrounding
individual comparisons motivates the present study.

We approach the problem through hierarchical Bayesian modelling,
which allows information to be shared across players while
retaining uncertainty about their individual conversion
deviations. Rather than comparing isolated point predictions,
we evaluate both shooters on identical contexts within each
joint posterior draw. Differences in their estimated conversion
probabilities can then be accumulated across a specified set
of opportunities, preserving the dependence introduced by
shared model parameters and player effects. The resulting
posterior distribution provides not only an estimate of the
expected-goal difference but also an assessment of its
uncertainty and the probability that either shooter would
have the higher expected conversion total.

Our analysis considers three related modelling approaches.
A contextual benchmark represents conversion using recorded
opportunity characteristics, while the primary hierarchical
model introduces a player-specific intercept to capture
conditional differences between shooters. We also examine a
provisional extension in which player deviations may vary
with shooting distance, angle and pressure. This additional
flexibility allows the relationship between shooter and
opportunity to change across contexts, although estimating
such variation reliably requires sufficiently informative
individual shooting records. The emphasis of the study is
therefore not on proposing a new class of predictive model,
but on developing a coherent framework for comparing players
while recognising the uncertainty and limitations of the
available evidence.

We investigate the framework using 35,700 eligible shots
from four leagues in the 2015/16 season. Chronological
validation examines how the models predict later observed
outcomes, while controlled experiments with known conversion
probabilities assess how effectively player differences can
be recovered as the available information increases.
Applications to Giroud--Vardy and Domenico Berardi--Nicola
Sansone then illustrate how the comparisons behave in
historical football settings, with particular attention
to the opportunities on which each player was observed.
Together, these analyses examine both the predictive
usefulness of the models and the strength of the evidence
supporting player-level distinctions.

Throughout this paper, we use the term ``counterfactual''
in a deliberately restricted sense. It refers to evaluating
an alternative shooter under the measured circumstances of
a recorded attempt, rather than predicting what would
actually happen following a substitution, transfer or
change in team selection. Replacing a player could alter
movement, interactions with teammates, shot selection and
the opportunities that arise, none of which is represented
by holding the recorded context fixed. Our objective is
instead to determine what can reasonably be inferred about
conditional conversion when two shooters are evaluated
on the same opportunities, and to make the limits of
those inferences as transparent as the estimates themselves.

\section{Positioning the football comparison}
\label{sec:related}

The development of expected goals modelling has been accompanied
by increasingly sophisticated representations of the circumstances
surrounding a shot. Early descriptions based primarily on shooting
location have been extended through strategic features derived
from spatiotemporal records \citep{code_7}, the integration of
positional and event information \citep{anzer_bauer_2021}, and
the incorporation of events preceding the attempt \citep{code_10}.
These developments belong to a wider body of research examining
how machine learning can contribute to the analysis of attacking
play \citep{code_1} and improve the predictive and analytical
usefulness of expected goals \citep{code_8}. Alongside improvements
in prediction, explanatory approaches have sought to make model
behaviour more interpretable through aggregated feature profiles
\citep{code_9}. Collectively, this literature has enriched our
understanding of how shooting opportunities can be represented
and evaluated. Yet describing an opportunity more accurately
does not necessarily establish how confidently we can distinguish
between the players who might attempt it.

Accounting for differences between shooters introduces a related
but distinct modelling challenge. Player- and position-adjusted
machine learning approaches \citep{code_3}, together with
Bayesian hierarchical corrections \citep{code_2}, allow conversion
probabilities to depart from a common expectation determined by
the recorded circumstances. The underlying principle of sharing
information across individuals while retaining uncertainty about
their differences is well established in Bayesian sports modelling
\citep{code_11}. Related applications include the modelling of
football team scores \citep{code_12} and individual performance
ratings \citep{code_13}, although these address different
outcomes from the shot-level conversion probabilities considered
here. Our work therefore does not seek to introduce player
adjustment or hierarchical pooling as new methodological ideas.
Instead, it examines how these established approaches can support
a more explicit assessment of uncertainty when comparing two
shooters on the same opportunities.

This distinction becomes important when moving from estimated
player effects to statements about individual comparisons.
Because predictions for different shooters depend on shared
contextual parameters, their uncertainties are not independent.
Consequently, examining separate marginal credible intervals
does not provide the uncertainty associated with the difference
between their predictions. Nor does ranking posterior means of
player intercepts fully address the question, since the resulting
difference in conversion probability depends on the opportunities
being evaluated and the nonlinear relationship between log-odds
and probabilities. By retaining the joint posterior structure
and evaluating both shooters within the same draws, paired
comparisons account for these dependencies at the level of
individual attempts and across collections of opportunities.

The interpretation of such comparisons also depends on what
is meant by a counterfactual change. In other sporting contexts,
virtual baseball simulations have been used to examine
alternative hitting strategies \citep{code_19}, while TacticAI
explores generated configurations of football corner kicks
\citep{code_22}. These applications consider changes to the
circumstances or decisions that shape an event. Our question
is deliberately narrower: given the circumstances of an
opportunity already recorded, how would the estimated conversion
probability differ if another shooter were evaluated in the
same situation? This distinction matters because the observed
data cannot establish how a substitution would alter the
opportunities created, the decisions taken or the subsequent
development of play. Drawing causal conclusions would require
additional assumptions about the observational evidence
\citep{code_15,code_16}, including careful consideration of
the estimand and the overlap between observed circumstances,
as emphasised in research on football crossing
\citep{code_21}. The contribution of the present study
therefore lies in connecting contextual prediction with
uncertainty-aware player comparison, while maintaining a
clear boundary between conditional conversion estimates
and the consequences of an actual personnel change.

\section{Data and experimental design}
\label{sec:data}

\subsection{Historical benchmark and eligibility}

Our analysis draws on StatsBomb's public event repository
\citep{code_24}, using revision
\texttt{4b73468fc5b0f1950f9f66fada70ad3a4f9327cb}.
The dataset covers the 2015/16 Premier League, La Liga,
Ligue 1 and Serie A seasons, providing a common historical
benchmark of event records and shot freeze frames across
four major European leagues. We use this population to
examine both predictive performance and player-level
comparisons, reserving later matches for chronological
validation. The resulting inferences are specific to the
observed season and population; neither performance in
other seasons nor the present-day ability of the players
is evaluated.

The original records contain 37,888 shots, of which
35,700 satisfy our eligibility criteria. We retain
source-classified open-play, non-penalty attempts with
valid shot and outcome information, together with
sufficient freeze-frame and goalkeeper geometry to
construct the contextual features. Importantly, players
with relatively few attempts remain in the dataset,
allowing us to examine how limited individual shooting
records affect the uncertainty of player comparisons.
Table~\ref{tab:data} summarises the resulting population
and its chronological partition. Exclusion categories
are mutually exclusive: where several criteria apply
to the same attempt, their combination is counted only
once. In particular, 354 penalties also lack freeze-frame
and goalkeeper geometry, so treating these reasons
independently would overstate the number of excluded shots.

\begin{table}[tb]
\caption{
\textbf{Shot eligibility and chronological partition.}
The initial four-league dataset contains 37,888 recorded
shots, of which 35,700 satisfy the eligibility criteria.
Exclusion categories are mutually exclusive and account
for 2,188 attempts in total, including cases where
multiple exclusion reasons coincide. The eligible
population is divided into training and validation
partitions according to match date.
}
\label{tab:data}
\centering\small
\begin{tabularx}{\linewidth}{Xr}
\toprule
Population or exclusion combination & Shots \\
\midrule
Raw four-league shot universe & 37,888 \\
Non-open-play free kick & 1,749 \\
Penalty + missing freeze frame + missing goalkeeper geometry & 354 \\
Penalty only & 46 \\
Missing goalkeeper geometry only & 32 \\
Non-open-play corner only & 6 \\
Corner + invalid coordinates & 1 \\
\midrule
Eligible primary population & 35,700 \\
Training partition & 25,253 \\
Validation partition & 10,447 \\
\bottomrule
\end{tabularx}
\end{table}

Shooters are identified using StatsBomb player identifiers,
while event identifiers uniquely define the attempts
included in each opportunity set. Player names are not
used to establish links with external ratings or
performance records. The four players considered in the
football applications are Jamie Vardy (10960, Leicester
City), Olivier Giroud (3604, Arsenal), Domenico Berardi
(7131, Sassuolo) and Nicola Sansone (6859, Sassuolo).

Using a fixed season, repository revision and set of
eligibility criteria ensures that predictive validation
and player comparison are grounded in the same underlying
population. It also places clear boundaries on the
interpretation of the findings. Changes in playing roles,
data annotation practices or the composition of shooting
opportunities may affect whether conclusions from this
historical benchmark remain applicable elsewhere.
Retaining players with limited shooting records is
particularly useful in this setting, since the amount
of available individual evidence is itself an important
part of the inferential problem.

\subsection{The geometry and circumstances of a shooting opportunity}
\label{sec:context}

Each shooting opportunity is represented by a
17-column design matrix combining geometric, situational
and execution-related information. Five numerical
variables describe shot distance, goal-opening angle,
goalkeeper distance, whether the attempt originates
inside the penalty area, and the number of defenders
within the shot triangle. These are accompanied by
three presence-encoded indicators for first-time shots,
pressure and one-on-one situations, together with
three body-part indicators and six technique indicators.
Head and Backheel serve as the respective reference
categories for body part and technique.

All numerical variables, including the penalty-area
indicator, are standardised using means and scales
estimated exclusively from the training data.
Categorical levels, reference categories and column
ordering are likewise established from training,
ensuring a consistent representation during validation
and player comparison. The context-dependent model
extension uses standardised distance and angle,
alongside the binary pressure indicator, from this
shared representation.

An important feature of the source data is that
situational annotations do not consistently distinguish
explicitly recorded false values from missing
annotations. We preserve these raw states before
encoding annotated true values as one and false or
absent values as zero. Consequently, a pressure
indicator of zero means that pressure was not recorded,
rather than demonstrating that the shooter was
unpressured. This distinction also applies when
interpreting comparisons restricted to particular
pressure categories.

\paragraph{Distance and opening to the goal.}

The distance between a shot and the goal centre is
an important characteristic of opportunity difficulty,
but proximity alone does not describe the opening
available to the shooter. An attempt taken near the
byline, for example, may be relatively close to the
goal while presenting a narrow angle between the
posts. We therefore represent these complementary
aspects of shooting geometry separately.

Let the shot location be $q=(q_x,q_y)$, the goal
centre $g=(120,40)$, and the goalposts
$\ell=(120,36)$ and $r=(120,44)$ in the source
coordinate system. Distance and goal-opening angle
are defined as

\begin{equation}
d(q)=\|q-g\|_2,\qquad
a(q)=\arccos\!\left\{
\frac{(\ell-q)^{\mathsf T}(r-q)}
{\|\ell-q\|_2\|r-q\|_2}
\right\}.
\label{eq:geometry}
\end{equation}

The angle measures the opening subtended by the
goal mouth from the shooting location, rather than
the direction of the shot or the orientation of
the player's body. Two attempts taken at similar
distances may therefore present substantially
different geometric opportunities depending on
their lateral positions. Although these features
capture important aspects of shooting geometry,
they do not describe ball speed, shot placement
or the difficulty of the preceding pass.

Distances are measured in \textbf{StatsBomb coordinate
units} on a $120\times80$ grid, rather than physical
metres, while angles are expressed in radians.
Pitch visualisations preserve an isotropic scale,
with all attempts oriented towards $x=120$.
The fitted models use the geometric quantities
defined in Equation~\eqref{eq:geometry} without
converting them into physical pitch dimensions.

\paragraph{Goalkeeper and defenders.}

Shot freeze frames provide the recorded positions
of players at the moment of an attempt
\citep{code_5}. From these observations, we derive
two additional descriptions of the defensive
configuration.

The goalkeeper feature measures the distance
between the opposing goalkeeper and the goal
centre. It does not represent the distance from
the shooter, the goalkeeper's coverage of the
goal or the direction of advancement. Consequently,
equal displacements to the side and towards the
field produce the same value.

\begin{figure}[tb]
\centering
\includegraphics[width=\linewidth]{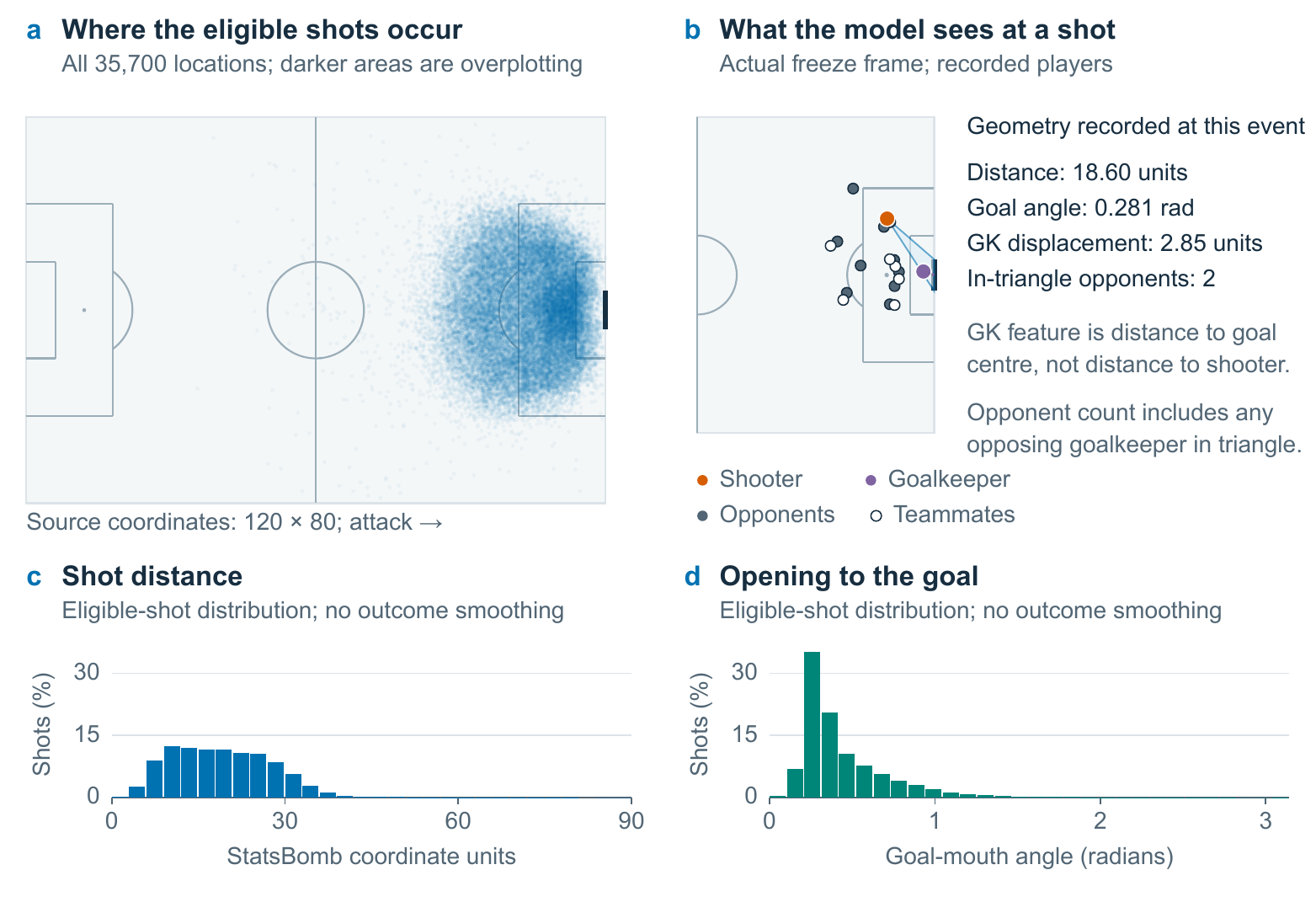}
\caption{
\textbf{Geometric characteristics of the recorded shooting opportunities.}
(a) Spatial distribution of all 35,700 eligible shots.
Darker regions indicate overlapping observations
rather than a smoothed probability surface.
(b) The first attempt in Vardy's opportunity set,
recorded on 8 August 2015, illustrating the triangle
formed by the shot location and goalposts, together
with the positions of the goalkeeper and other
recorded players. Goalkeeper distance is measured
from the goal centre, while the defensive count
includes opposing players strictly inside the
triangle, including the goalkeeper when applicable.
(c,d) Distributions of shot distance and goal-opening
angle, each displayed using 30 equally spaced bins.
Pitch coordinates follow the original
$120\times80$ system, with attacks directed
towards $x=120$. Distances are expressed in
source coordinate units and angles in radians.
These visualisations describe the geometry of
observed opportunities rather than differences
in player ability.
}
\label{fig:difficulty}
\end{figure}

We also count opposing players positioned strictly
inside the triangle formed by the shot location
and the two goalposts. The goalkeeper contributes
to this count whenever positioned within the
triangle. Together, these variables provide
a compact description of the defensive arrangement
visible in the freeze frame, although they cannot
fully represent potential blocking actions,
player movement or the subsequent trajectory
of the ball. Figure~\ref{fig:difficulty} illustrates
these geometric relationships using an observed
shooting opportunity.

\paragraph{Pressure, execution and measurement.}

The interpretation of situational features requires
particular care because their annotations reflect
both the circumstances of an attempt and the
conventions used to record them. Pressure, for
example, may describe a constraint on shooting
execution, but its recorded presence may also
depend on the situations in which a player
chooses or is able to attempt a shot.

Annotation conventions can differ between
data providers \citep{code_6}. We therefore
interpret pressure according to the source
definition and the presence-only encoding
described above. First-time and one-on-one
indicators are subject to the same distinction
between recorded presence and unrecorded absence.

Body part and shooting technique provide
additional information about execution, but
neither captures the complete sequence of
actions leading to a shot. Accordingly,
estimated coefficients and subgroup comparisons
describe associations within the measured
circumstances. They should not be interpreted
as establishing what would happen if a
defender were removed or a particular
constraint on execution were changed.

\subsection{Chronological partition and evaluation roles}

We divide the eligible observations chronologically,
assigning shots recorded up to and including
28 February 2016 to training and those from
29 February 2016 onwards to validation.
The split is performed at match level, ensuring
that the two partitions share no matches,
events or calendar dates.

The training partition contains 25,253 shots,
including 2,378 goals from 1,699 observed
players. The validation partition comprises
10,447 shots and 1,064 goals across 445 matches,
corresponding to goal rates of 9.417\% and
10.185\%, respectively. Of the validation
observations, 247 shots and 16 goals belong
to 100 players who do not appear in training.
All three modelling approaches are evaluated
using the same validation observations, allowing
their predictive performance to be compared
on a common basis.

The chronological split serves a different
purpose from the player-level comparisons.
Predictive evaluation uses the observed
outcomes of later shots to assess how well
models fitted on earlier data generalise
to subsequent matches. Player comparisons,
by contrast, evaluate alternative shooters
on each factual shooter's complete eligible
set of opportunities, while distinguishing
between training-period and validation-period
contexts.

Observed goals remain part of the factual
record and do not update the training
posterior during these comparisons.
Applications involving training-period
opportunities are therefore in-sample,
whereas those involving validation-period
opportunities examine later contexts
without refitting the models. Neither
setting reveals the conversion probability
that an alternative player would actually
have had. We address this remaining
inferential difficulty separately through
controlled experiments in which the
underlying conversion probabilities
are known.

\section{Hierarchical inference: separating opportunity and shooter}
\label{sec:models}

\subsection{Model specifications and priors}

We consider three models that differ in how they represent
the contribution of the shooter to the probability of
scoring. The contextual benchmark, M0, estimates conversion
probabilities from the recorded circumstances of each
attempt. The primary hierarchical model, M1, introduces
a player-specific intercept, allowing shooters to differ
after accounting for those circumstances. A provisional
extension, M2, permits these player differences to vary
with selected characteristics of the opportunity.

For shot $j$ taken by player $i$, let $y_{ij}$ indicate
whether the attempt resulted in a goal, and let
$x_{ij}\in\mathbb{R}^{17}$ denote its contextual features.
All three models share a Bernoulli likelihood,

\begin{equation}
y_{ij}\sim\operatorname{Bernoulli}(p_{ij}),\qquad
p_{ij}=\operatorname{logistic}(\eta_{ij}).
\label{eq:likelihood}
\end{equation}

The contextual benchmark and primary hierarchical model
are specified as

\begin{align}
\mathrm{M0}:\quad\eta_{ij}
&=\alpha+x_{ij}^{\mathsf T}\beta,
\label{eq:m0}\\
\mathrm{M1}:\quad\eta_{ij}
&=\alpha+x_{ij}^{\mathsf T}\beta+u_i,
\quad
u_i\mid\sigma_u\sim\mathcal{N}(0,\sigma_u^2).
\label{eq:m1}
\end{align}

In M1, the player intercept represents a conditional
departure from the conversion expectation established
by the recorded context. It operates on the log-odds
scale and is not intended to distinguish individual
components of finishing ability, such as aerial
technique, shooting execution or tactical decision-making.

The provisional extension M2 introduces a further
degree of flexibility. Let $z_{ij}$ contain distance,
angle and pressure selected from the contextual vector.
The model then becomes

\begin{equation}
\mathrm{M2}:\quad
\eta_{ij}
=\alpha+x_{ij}^{\mathsf T}\beta+u_i
+z_{ij}^{\mathsf T}v_i,\qquad
v_{ik}\mid\sigma_{v,k}
\sim\mathcal{N}(0,\sigma_{v,k}^2).
\label{eq:m2}
\end{equation}

Unlike M1, which assigns each player a constant
deviation on the log-odds scale, M2 allows the
deviation to change with particular shooting
conditions. The intercept and three slope
hierarchies are specified independently, without
estimating covariance between their effects.
For computation, we use the non-centred
representations $u_i=\sigma_u\tilde u_i$ and
$v_{ik}=\sigma_{v,k}\tilde v_{ik}$, with independent
standard normal latent variables.

The primary prior specification is

\begin{align}
\alpha
&\sim\mathcal{N}(\logit(\bar y_{\mathrm{train}}),0.5^2),
&
\beta_l
&\sim\mathcal{N}(0,1),
\label{eq:priors}\\
\sigma_u
&\sim\operatorname{HalfNormal}(0.5),
&
\sigma_{v,k}
&\sim\operatorname{HalfNormal}(0.5).
\end{align}

Normal distribution arguments specify variances,
whereas HalfNormal arguments specify scales.
The applicable priors are shared across models,
with M0 omitting the player hierarchy. The
intercept prior is centred empirically using
only the training goal rate, corresponding to
a logit centre of approximately $-2.264$
in the real-data analysis. Neither validation
outcomes nor external player ratings contribute
to this specification.

To examine sensitivity to prior assumptions,
we additionally consider M0/M1 specifications
with $\alpha\sim\mathcal{N}(0,2.5^2)$,
$\beta_l\sim\mathcal{N}(0,1.5^2)$ and
$\sigma_u\sim\operatorname{HalfNormal}(1)$.
These analyses provide a supplementary assessment
of prior sensitivity while leaving the primary
specification unchanged.

\subsection{What does pooling mean for a football player?}
\label{sec:pooling}

Estimating differences between footballers presents
a familiar statistical difficulty: the available
evidence is distributed unevenly across individuals.
Some players contribute many attempts under varied
circumstances, whereas others have only a handful
of recorded shots. Treating every player independently
would allow sparse records to produce potentially
extreme estimates, even when the evidence supporting
those estimates is weak.

Hierarchical modelling addresses this problem
through partial pooling. In M1, each player has
an individual deviation $u_i$, but the distribution
of these deviations is estimated collectively
from the population. This allows information to
be shared across players, following a principle
widely established in Bayesian sports modelling
\citep{code_11}. When a player's shooting record
contains little information, the hierarchical
structure draws the estimated deviation towards
the population centre while retaining uncertainty
about both the individual effect and the population
distribution. As the evidence becomes more
informative, the individual estimate can depart
more substantially from that centre.

This behaviour discourages strong conclusions
based on a small number of successful or
unsuccessful finishes. It also introduces a
trade-off: although pooling can stabilise
estimates from sparse records, genuine but
small player differences may be attenuated.
Moreover, the information contained in a
shooting record depends not only on the
number of attempts, but also on their
outcomes and the circumstances in which
they occurred.

The implications of pooling become particularly
clear when comparing players across different
opportunities. Within any posterior draw of M1,
the ordering of shooters is determined by
their intercepts. Because the logistic function
is monotonic, the player with the larger
intercept has the higher estimated conversion
probability in every context. The magnitude
of that advantage, however, is not constant
on the probability scale. The same difference
in log-odds can translate into different
probability differences depending on the
underlying difficulty of the opportunity.

Consequently, two sets of shots may produce
different aggregate expected-goal contrasts
even when the relative ordering of the players
remains unchanged. Reversals in that ordering
across shooting conditions require a more
flexible representation, such as M2.

By introducing player-specific associations
with standardised distance, standardised
angle and binary pressure, M2 allows the
estimated advantage of one shooter over
another to vary across circumstances.
This flexibility may be useful when
individual shooting records contain
sufficiently diverse information, but
it also places greater demands on the
data and may increase uncertainty or
sensitivity to contextual exposure.
Furthermore, the independent hierarchies
do not estimate correlations between
intercept and slope deviations.

For these reasons, we treat M2 as a
supplementary specification rather than
the primary basis for player comparison.
Its real-data sampling qualification
is retained when interpreting the
football applications. In particular,
variation in modelled conversion
probabilities should not, by itself,
be regarded as evidence of a player's
specialisation in particular shooting
conditions.

No external player-specific information
enters the prior distributions. The
hierarchical effects are centred at
zero, expressing exchangeability
conditional on the measured context,
while the overall intercept is centred
using the training goal rate. This
assumption supports information sharing
without requiring players to occupy
identical roles or encounter equivalent
opportunities. The resulting effects
are interpreted as conditional
conversion deviations rather than
isolated measures of finishing ability,
subject to the selection and
measurement limitations discussed
in Section~\ref{sec:discussion}.

\subsection{Paired conditional comparison}

The central comparison in this study asks
how two players' estimated conversion
probabilities differ when evaluated
on exactly the same recorded opportunities.
We begin by specifying a fixed set of
contexts, denoted by
$\mathcal{C}=\{x_1,\ldots,x_N\}$.
For a factual shooter $A$ and an
alternative shooter $B$, the difference
in conversion probability for an
opportunity $x$ is evaluated within
posterior draw $s$ as

\begin{equation}
\Delta^{(s)}(A\to B,x)
=p_B^{(s)}(x)-p_A^{(s)}(x).
\label{eq:contrast}
\end{equation}

Both probabilities are calculated using
the same draw of the shared contextual
parameters. This pairing is essential:
the comparison concerns the difference
between shooters under a common set
of model conditions, rather than
differences between independently
sampled predictions.

For each posterior draw, we aggregate
the estimated probabilities across
the specified opportunities,

\begin{equation}
\begin{aligned}
G_A^{(s)}
&=\sum_{x\in\mathcal C}p_A^{(s)}(x),
\qquad
G_B^{(s)}
=\sum_{x\in\mathcal C}p_B^{(s)}(x),\\
D^{(s)}
&=\sum_{x\in\mathcal C}
\Delta^{(s)}(A\to B,x).
\end{aligned}
\label{eq:aggregate}
\end{equation}

The resulting distribution of $D^{(s)}$
describes uncertainty in the difference
between the players' expected-goal totals.
We summarise this distribution using
its posterior mean, an equal-tail
95\% credible interval, and the
posterior probabilities $\Prb(D>0)$
and $\Prb(D<0)$.

Importantly, these quantities describe
differences in expected conversion,
not the distribution of goals that
would be realised if the opportunities
were replayed. Simulating additional
Bernoulli outcomes would introduce
variation associated with goal
realisation, which is distinct
from uncertainty about the
underlying conversion probabilities.

The direction of the comparison
also requires careful interpretation.
When the context set remains
unchanged, exchanging shooters
reverses the sign of the contrast,
and comparing a player with
themselves gives zero. However,
evaluating Giroud on Vardy's
recorded opportunities is not
the numerical reverse of
evaluating Vardy on Giroud's
opportunities, because the
two calculations involve
different sets of contexts.

The comparison therefore has a
deliberately restricted interpretation.
It estimates how conversion
probabilities differ under
fixed measured circumstances,
without establishing whether
the alternative player would
have encountered those
circumstances in an actual match.
Where a player has little
experience of the evaluated
conditions, the calculation
also relies on extrapolation
beyond their observed exposure.
We examine this issue in
Section~\ref{sec:support}
and distinguish conditional
comparison from causal
substitution in
Section~\ref{sec:discussion}.

\subsection{Why opportunity-level uncertainty must remain paired}
\label{sec:pairing}

When comparisons involve many
opportunities, the uncertainty
associated with their aggregate
difference cannot be obtained
by treating each shot independently.
Although the attempts are distinct,
their estimated conversion
probabilities depend on shared
model parameters and player
effects. Uncertainty in those
quantities therefore influences
multiple predictions simultaneously.

For an aggregate contrast $D$,
the posterior variance can
be expressed as

\begin{equation}
\operatorname{Var}(D\mid\mathcal D)
=\sum_j\operatorname{Var}(\Delta_j\mid\mathcal D)
+2\sum_{j<k}
\operatorname{Cov}(\Delta_j,\Delta_k\mid\mathcal D).
\label{eq:covariance}
\end{equation}

The covariance terms capture
the dependence between
opportunity-level differences.
Ignoring them would generally
produce an incorrect assessment
of uncertainty in the total.

We preserve this dependence
by calculating every
opportunity-level contrast
within the same posterior
draw before aggregation.
The two shooters are likewise
evaluated using aligned
draws, so that shared
uncertainty is retained.
Independently shuffling
one player's posterior
draws would break this
relationship and alter
the distribution of the
comparison.

For the real-data
applications, we use
4,000 aligned evaluations
for each comparison,
maintaining a common
chain and draw index
across both shooters
and all opportunities.

The resulting credible
interval describes
uncertainty about the
difference in expected
goals, rather than
the variability of
realised goal totals.
Observed goals remain
part of the factual
record and contribute
to model estimation
or predictive evaluation
as appropriate. They
cannot reveal how many
goals an alternative
shooter would have
scored from those
same opportunities.

\subsection{Inference, prediction and support}

Posterior inference is performed
in PyMC \citep{pymc2023}
using the No-U-Turn Sampler
(NUTS) \citep{hoffman2014}.
We run four chains, each
with 1,000 warm-up iterations
and 1,000 retained posterior
draws. Sampling uses a target
acceptance probability of
0.95, a maximum tree depth
of 10 and random seed 2026.

The real-data computational
environment comprises
Python 3.14.8, PyMC 6.3.2,
ArviZ 1.3.0 and
PyTensor 3.3.3.

We assess sampling quality
using rank-based
$\widehat R$, bulk and
tail effective sample sizes
(ESS), divergent transitions,
energy-transition
Bayesian fraction of
missing information
(BFMI), and tree-depth
saturation
\citep{arviz2019,vehtari2021}.
Our adequacy criteria require
$\widehat R\leq1.01$,
ESS of at least 400,
no divergences or
tree-depth saturation,
BFMI of at least 0.3,
and finite diagnostics
for all relevant parameters.

These diagnostics assess
the numerical reliability
of posterior sampling,
but they do not establish
that the model adequately
represents the data.
In particular, the
absence of divergent
transitions is not
sufficient evidence
of satisfactory inference.
Bayesian model assessment
also requires checks
beyond numerical convergence
\citep{gabry2019}.

Prediction depends on
whether the player was
observed during training.
For previously observed
players, we use their
fitted posterior effects.
For an unseen player,
each posterior sample
includes a population
effect shared across
that player's opportunities.
In M2, this includes
both the intercept
and slope effects.

This approach integrates
over the population
distribution while
preserving dependence
between predictions
for the same unseen
player. It represents
uncertainty about the
player's effect without
claiming to identify
their individual
conversion deviation.
All four players in
the football applications
appear in training,
so their comparisons
use fitted player
effects throughout.

We interpret the
resulting comparisons
alongside the players'
observed contextual
exposure. This includes
training shot counts,
pressure annotations,
and the relationship
between evaluation
contexts and each
player's recorded
distance, angle and
pressure ranges.

These summaries help
identify comparisons
that rely on unfamiliar
shooting circumstances,
which may not be
apparent from posterior
intervals alone.
They do not, however,
establish joint
contextual overlap
or provide calibrated
categories of
predictive reliability.

An OpenAI language-model
assistant supported
manuscript organisation
and drafting from
the study materials.
Responsibility for
verifying the text,
visualisations and
scientific content
rests with the authors.

\section{Can the model predict later football outcomes?}
\label{sec:results}

Before examining differences between individual shooters, we
first consider whether the models provide reliable predictions
for opportunities encountered later in the season. This
evaluation serves two purposes. It establishes whether the
recorded shooting circumstances contain useful information
about goal probability, and it examines whether accounting
for individual players improves predictions beyond what
can already be explained by those circumstances. Importantly,
neither result directly establishes whether the model can
distinguish two shooters on an identical opportunity.

Predictive performance is evaluated against the observed
binary outcomes of validation-period shots using log loss
and Brier score. Both are proper scoring rules, making them
appropriate for assessing probabilistic forecasts rather
than simply classifying attempts as goals or misses
\citep{waghmare2026}. We compare the fitted models with
a constant predictor based on the training goal rate.
Calibration is examined using ten equal-frequency bins,
where agreement between predicted probabilities and
observed goal frequencies indicates reliable probability
estimation \citep{vancalster2019}. To quantify uncertainty
in differences between predictive scores, we use 4,000
paired bootstrap resamples of the 445 validation matches.
Resampling at match level preserves dependence between
attempts from the same match, although the resulting
intervals are conditional on posterior-mean predictions
and do not incorporate posterior parameter uncertainty.

\begin{table}[tb]
\caption{
\textbf{Predictive performance and posterior sampling diagnostics.}
The upper panel reports chronological validation scores
for the constant training-rate benchmark and the three
fitted models. Lower log loss and Brier score indicate
better probabilistic predictions. The lower panel
summarises posterior sampling diagnostics for M0,
the primary hierarchical model M1, and the provisional
context-dependent extension M2. BFMI is reported as
the minimum across chains, while effective sample
sizes refer to the reported diagnostic summaries.
No model exhibits divergent transitions or tree-depth
saturation. M2 does not satisfy the specified
$\widehat R$ criterion for every player-slope
parameter and is therefore treated as provisional.
}
\label{tab:validation}
\centering\small
\begin{tabular}{lrr}
\toprule
Predictor & Log loss & Brier score \\
\midrule
Training-rate constant & 0.32946 & 0.0915 \\
M0 & 0.26687 & 0.0770 \\
M1 & 0.26677 & 0.0770 \\
M2 (provisional) & 0.26676 & 0.0770 \\
\bottomrule
\end{tabular}
\par\medskip
\setlength{\tabcolsep}{4pt}
\begin{tabular}{lrrrrr}
\toprule
Model & Max $\widehat R$ & Bulk ESS & Tail ESS & Divergences & BFMI \\
\midrule
M0 & 1.0055 & 1,282 & 1,717 & 0 & 0.99 \\
M1 & 1.0081 & 695 & 1,362 & 0 & 0.86 \\
M2 & 1.0128 & 601 & 1,112 & 0 & 0.92 \\
\bottomrule
\end{tabular}
\end{table}

The results reveal a clear difference between the value of
contextual information and the additional contribution of
player-specific effects. All three fitted models substantially
outperform the constant-rate benchmark, reducing log loss
from 0.32946 to approximately 0.2668 and Brier score from
0.0915 to 0.0770 (Table~\ref{tab:validation}). These
improvements demonstrate that the recorded circumstances
of an attempt provide considerably more information about
its outcome than the overall scoring frequency alone.
However, the predictive differences among M0, M1 and M2
are much smaller, suggesting that the additional player
effects contribute relatively little to average predictive
performance on the validation population.

The comparison between M1 and M0 illustrates this
distinction particularly clearly. Introducing the
hierarchical player intercept reduces log loss by
0.000103, with a paired match-bootstrap 95\% interval
of $[-0.000198,-0.000009]$. Although this interval
excludes zero under the specified resampling procedure,
the improvement is extremely small in magnitude.
Consequently, the evidence for a detectable difference
in predictive scores should not be confused with
evidence of a practically important improvement.
Allowing player effects to vary with shooting conditions
produces even less additional benefit. The M2--M1
log-loss difference is $-0.000012$, with a 95\%
interval of $[-0.00016,+0.00014]$, providing no clear
predictive distinction between the two hierarchical
specifications.

\begin{figure}[tb]
\centering
\includegraphics[width=\linewidth]{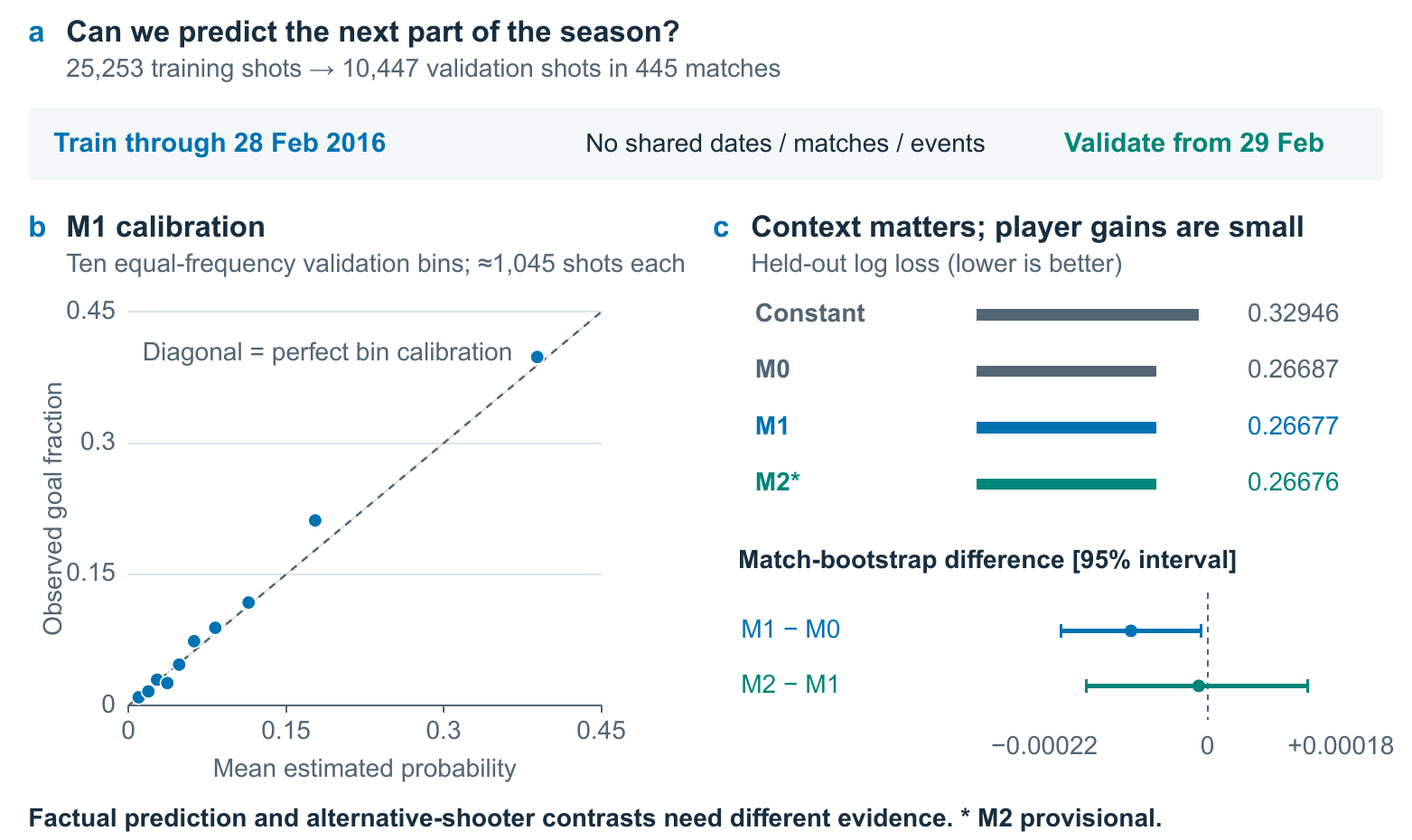}
\caption{
\textbf{Chronological predictive evaluation and uncertainty in model comparisons.}
(a) Training and validation periods, separated
chronologically without shared matches, events
or calendar dates.
(b) Calibration of M1 across ten equal-frequency
validation bins. Points represent mean predicted
probabilities and observed goal frequencies
within each bin, while the diagonal indicates
perfect calibration. The plotted points do not
include independently replicated uncertainty
intervals.
(c) Log loss and paired match-bootstrap 95\%
intervals for differences between posterior-mean
predictors. All fitted models improve upon the
constant training-rate benchmark, but differences
among M0, M1 and provisional M2 remain small.
Bootstrap intervals account for match-level
resampling without incorporating posterior
parameter uncertainty. These evaluations concern
predictions for observed outcomes, rather than
the recovery of unobserved alternative-shooter
conversion differences.
}
\label{fig:prediction}
\end{figure}

The models also exhibit generally satisfactory
calibration. The calibration slopes are close to
unity, reaching 1.006 for M0 and 1.007 for M1,
although both show a modest tendency to underestimate
the observed scoring frequency. For M1, the mean
predicted goal probability is approximately 9.67\%,
compared with an observed validation rate of 10.18\%.
Its equal-frequency calibration error is approximately
0.008. Figure~\ref{fig:prediction} places these
results alongside the chronological partition and
the uncertainty associated with predictive score
differences. The overall picture is consistent:
contextual modelling improves probability estimation,
while increasingly elaborate representations of
player differences offer little additional benefit
on the observed validation outcomes.

Predictive performance must nevertheless be considered
alongside the reliability of posterior inference.
Both M0 and M1 satisfy the prespecified sampling
criteria, whereas M2 exhibits a limited number of
convergence concerns. Specifically, 14 of its 5,097
player-slope entries exceed the
$\widehat R=1.01$ threshold, although none exceeds
1.05 and all effective sample sizes remain at
least 400. These findings do not invalidate the
entire model, but they limit the confidence with
which its player-specific slope estimates can
be interpreted. We therefore retain M2 as a
provisional extension and base the principal
player comparisons on M1.

The hierarchical estimates themselves provide
further insight into the limited contribution
of player effects. In M1, the posterior mean
of the between-player logit standard deviation
is 0.088, with an approximate 95\% interval
of $[0.004,0.205]$. The concentration of
this distribution towards relatively small
values suggests that differences between
players are only weakly identified beyond
the recorded shooting circumstances.
Under the supplementary prior specification,
the corresponding posterior mean is 0.093,
with similar predictive performance.
This agreement indicates that the findings
are reasonably consistent across the two
examined specifications, although it does
not establish robustness to alternative
prior choices more generally. No equivalent
prior-sensitivity analysis was performed
for M2.

An additional limitation concerns predictions
for players who were absent from the training
population. The validation set contains
247 attempts and only 16 goals from such
players. M1 achieves a log loss of
approximately 0.1891 on these observations,
but the small number of realised goals
makes this result difficult to interpret
as evidence of reliable generalisation
to previously unseen shooters. In particular,
the apparent quality of an average
probability forecast cannot establish
whether individual player effects have
been estimated accurately.

Taken together, the validation results
demonstrate that the models can predict
observed shooting outcomes substantially
better than a constant-rate benchmark,
but they offer much weaker evidence
about the value of distinguishing
individual shooters. A player adjustment
may improve an average predictive score
without providing enough information
to resolve the difference between two
particular players. Conversely, a model
that predicts well across thousands
of attempts may still be highly
uncertain about the relative conversion
probabilities of those players when
evaluated on the same opportunity.
Predictive accuracy and the identification
of individual differences are therefore
related, but fundamentally distinct,
questions.

This distinction is especially important
for the historical applications that follow.
All player comparisons use the posterior
estimated from training-period outcomes.
Opportunities recorded during training
contributed to model estimation, whereas
validation-period contexts are evaluated
without updating the posterior. Comparisons
over a player's complete seasonal
opportunity set consequently combine
in-sample and later contexts, which are
distinguished in the corresponding
visualisations. Neither setting constitutes
a held-out validation of the alternative-
shooter comparison, because the outcome
that would have been observed for the
alternative player remains unavailable.
To investigate whether such differences
can be recovered when their underlying
values are known, we next turn to
controlled experiments.

\section{Known-truth football experiments: when is a contrast identifiable?}
\label{sec:recovery}

\subsection{Fixed situations with known conversion differences}

The historical data allow us to evaluate predictions against observed
goals, but they cannot reveal what would have happened had a different
player attempted the same shot. Consequently, the accuracy of an
alternative-shooter comparison cannot be established directly from
observational outcomes. To examine this problem under conditions where
the underlying differences are known, we conduct controlled experiments
in which player conversion probabilities are specified before model
fitting. This allows us to distinguish uncertainty arising from limited
information from discrepancies between the estimated and generating
conversion probabilities.

The experiments retain the contextual definitions used in the historical
analysis, while generating new player identities and Bernoulli outcomes
from a known logistic mechanism. Contextual observations are sampled
jointly from the training covariate pool, preserving the relationships
among recorded shooting characteristics without reusing actual goal
outcomes. Player heterogeneity is introduced through intercepts and
three context-dependent slopes, with generating scales of 0.20 for
intercepts and $(0.15,0.15,0.20)$ for distance, angle and pressure,
respectively. The realised effect columns are centred across simulated
players. The generating intercept is calibrated once against a fixed
reference mechanism near the observed training goal prevalence, rather
than adjusted separately to obtain favourable results in individual
experimental conditions.

We begin with six scenarios designed to examine different sources of
difficulty in recovering player differences. S0 represents a population
without player heterogeneity, while S1 introduces player intercepts
and S2 allows both intercept and contextual slope differences. The
remaining scenarios examine departures from the more regular conditions
of S2: S3 introduces unequal shooting volumes based on empirical
records, S4 assigns players to separated regions of the shooting-distance
distribution, and S5 introduces a nonlinear distance effect omitted
from the fitted M2 specification. Each scenario changes its designated
component while retaining the common seed and evaluation contexts.

The generating mechanism is defined by

\begin{equation}
p_i^{\mathrm{true}}(x)
=
\operatorname{logistic}\!\left(
\alpha_*
+x^{\mathsf T}\beta_*
+u_{i,*}
+z(x)^{\mathsf T}v_{i,*}
+\gamma[d_{\mathrm{std}}(x)^2-m_2]
\right).
\label{eq:synthetic}
\end{equation}

Here, starred quantities denote generating parameters, and $m_2$
is the mean squared standardised distance in the training covariate
pool. The nonlinear coefficient is set to $\gamma=0.25$ in S5 and
zero elsewhere, while player intercepts and slopes are disabled
as required by S0 and S1. Because the generating probabilities
are known, the true difference between players $A$ and $B$ can
be calculated directly as
$p_B^{\mathrm{true}}(x)-p_A^{\mathrm{true}}(x)$.
The corresponding aggregate truth is obtained by summing these
differences over the fixed evaluation opportunities, independently
of fitted predictions or realised goal totals. Although contrasts
can be reconstructed in both directions, recovery is evaluated
using canonical $A<B$ pairs to avoid duplicating comparisons
with opposite signs and cancelling signed biases.

The initial experiment comprises 18 model fits, covering the six
scenarios under M0, M1 and M2. Fifteen satisfy the prespecified
sampling criteria, while M2 exhibits narrow $\widehat R$
exceedances in S0, S2 and S3. Across the pilot, every 95\%
credible interval for both individual-context and aggregate
contrasts includes zero. The models therefore provide little
directional resolution under these conditions, even though
the generating differences are known. This finding should not
be interpreted as evidence that the players are equivalent
or that heterogeneity is absent. Rather, it illustrates how
limited information can prevent the available data from
supporting confident distinctions between genuinely different
conversion probabilities.

\subsection{Increasing observations without changing the football truth}

The initial experiment identifies several potential obstacles
to recovering player differences, but it does not isolate the
effect of having more observations for each player. We therefore
conduct a separate information experiment in which the
underlying conversion mechanism remains fixed while the
amount of training evidence increases. Twenty simulated
players, their generating abilities, the model coefficients
and 32 evaluation opportunities are held constant throughout.
For each player, we generate 300 shots once and construct
nested training datasets containing the first 15, 60 and
300 observations. The smaller datasets are thus contained
within the larger ones, preserving their original contexts
and outcomes.

This design allows us to examine how additional observations
affect inference without changing the players being compared
or the opportunities on which their differences are evaluated.
We fit M1 to the intercept-only S1 mechanism and M2 to the
context-dependent S2 mechanism. Two unseen player identities
provide secondary controls, although the principal analysis
concerns the 190 canonical pairs formed by the 20 observed
players. The 15-shot samples are generated separately from
those in the initial pilot, so the two experiments should
not be treated as successive stages of a single dataset.

Each synthetic fit uses four chains with 500 warm-up iterations
and 500 retained posterior draws, applying the primary prior
specification, a target acceptance probability of 0.95 and
random seed 2026. Where the intercept prior is centred
empirically, its centre is calculated only from the training
outcomes available in the corresponding subset.

Because the true conversion probabilities are known, recovery
can be assessed directly rather than inferred from predictive
performance alone. We examine posterior-mean bias and
root-mean-square error (RMSE), the coverage and width of
equal-tail 90\% and 95\% credible intervals, directional
accuracy, the resolution of meaningful differences and
aggregate expected-goal recovery. These measures capture
different aspects of inference: an estimate may approach
the truth without becoming sufficiently precise to
distinguish a difference from zero.

For the primary analysis, we define a meaningful per-shot
contrast as one whose absolute true probability difference
exceeds the prespecified pilot tolerance $\epsilon=0.01$.
Sensitivity analyses consider tolerances of 0.005 and 0.02.
A meaningful contrast is correctly resolved when its
credible interval excludes zero in the true direction,
incorrectly resolved when the interval excludes zero in
the opposite direction, and unresolved when the interval
includes zero. We also record false directional confidence,
which includes both wrong-direction intervals and intervals
excluding zero for effects within $\pm\epsilon$, even when
the estimated direction is correct. The tolerance is an
experimental convention for distinguishing small effects,
not an empirically validated threshold for football decisions.

\begin{table}[tb]
\caption{
\textbf{Recovery of known player differences as training information increases.}
Results concern canonical seen--seen comparisons at the primary
tolerance $\epsilon=0.01$, with 15, 60 or 300 training shots
per player. RMSE measures posterior-mean error, while
Width$_{95}$ is the mean width of equal-tail 95\% credible
intervals; both are expressed as probability differences.
Cov$_{95}$ and False$_{95}$ are percentages calculated across
6,080 dependent pair/context cells. Correct$_{95}$ counts
meaningful contrasts resolved in the true direction, among
1,763 meaningful cells in S1 and 2,299 in S2. False confidence
includes wrong-direction claims and confident claims about
effects within $\pm\epsilon$. The asterisk identifies the
provisional S2/M2 fit with 60 shots per player. Results
describe one simulated realisation rather than repeated
sampling coverage or validated football sample-size requirements.
}
\label{tab:recovery}
\centering
\scriptsize
\setlength{\tabcolsep}{3pt}
\begin{tabular}{llrrrrr}
\toprule
Scenario/model & Shots/player & RMSE & Width$_{95}$ &
Cov$_{95}$ & Correct$_{95}$ & False$_{95}$ \\
\midrule
S1/M1 & 15  & 0.0231 & 0.1564 & 99.79 & 0   & 0.000 \\
S1/M1 & 60  & 0.0201 & 0.1039 & 99.49 & 0   & 0.000 \\
S1/M1 & 300 & 0.0139 & 0.0499 & 98.27 & 0   & 0.000 \\
S2/M2 & 15  & 0.0528 & 0.3029 & 99.90 & 13  & 0.033 \\
S2/M2$^*$ & 60 & 0.0337 & 0.1589 & 99.74 & 33 & 0.016 \\
S2/M2 & 300 & 0.0224 & 0.0792 & 98.16 & 169 & 0.855 \\
\bottomrule
\end{tabular}
\end{table}

The results show a consistent improvement in estimation accuracy
and precision as more observations become available
(Table~\ref{tab:recovery}). In both scenarios, posterior-mean
RMSE decreases and credible intervals become narrower at each
successive training size. Under S1, RMSE falls from 0.0231
with 15 shots to 0.0139 with 300, while mean 95\% interval
width decreases from 0.1564 to 0.0499. The corresponding
reductions under S2 are from 0.0528 to 0.0224 for RMSE and
from 0.3029 to 0.0792 for interval width. Additional
observations therefore provide more accurate and concentrated
estimates in this controlled realisation.

Improved estimation does not, however, translate directly
into confident identification of meaningful differences.
With 300 shots per player, S1 achieves posterior-mean
directional accuracy of 81.7\%, yet none of its 1,763
meaningful contrasts is resolved at the 95\% level.
The estimates increasingly favour the correct direction,
but their uncertainty remains sufficient to include zero.
S2 achieves directional accuracy of 78.8\% at the same
training size and correctly resolves 169 of its 2,299
meaningful contrasts, leaving 2,130 unresolved. Thus,
even with substantially more information, most
player-level differences remain too uncertain to
support a confident directional conclusion.

A further distinction emerges when considering effects
that are smaller than the declared tolerance. At 300
shots per player, S2 produces no wrongly resolved
meaningful contrasts at the 95\% level, but 52 confident
claims concern near-null effects. Of these, 50 have
the correct direction and two have the wrong direction.
Together, they represent 0.855\% of all evaluated cells
and 23.5\% of the 221 confident claims. These findings
illustrate why excluding zero is not sufficient to
establish that an estimated advantage is meaningful.
Greater precision can make a small effect statistically
distinguishable from zero without making it important
under the chosen interpretation threshold.

The same general pattern appears when differences are
aggregated across the 32 fixed evaluation opportunities.
Aggregate expected-goal RMSE decreases from 0.5459 to
0.3205 in S1 and from 0.7708 to 0.3903 in S2 as
training size increases from 15 to 300 shots. Nevertheless,
at the largest training size, none of the 83 meaningful
aggregate contrasts is correctly resolved under S1,
compared with 15 of 110 under S2. These aggregate
intervals are constructed by summing contrasts within
posterior draws, preserving their shared uncertainty
rather than treating opportunity-level differences
as independent.

All six fits in the information experiment complete
without divergent transitions or tree-depth saturation.
The S2 fit with 60 shots per player remains provisional,
however, because one non-centred slope parameter has
$\widehat R=1.01199$, marginally exceeding the specified
criterion. This qualification concerns sampling
adequacy and should be distinguished from the broader
uncertainty in recovering player differences.

\begin{figure}[tb]
\centering
\includegraphics[width=\linewidth]{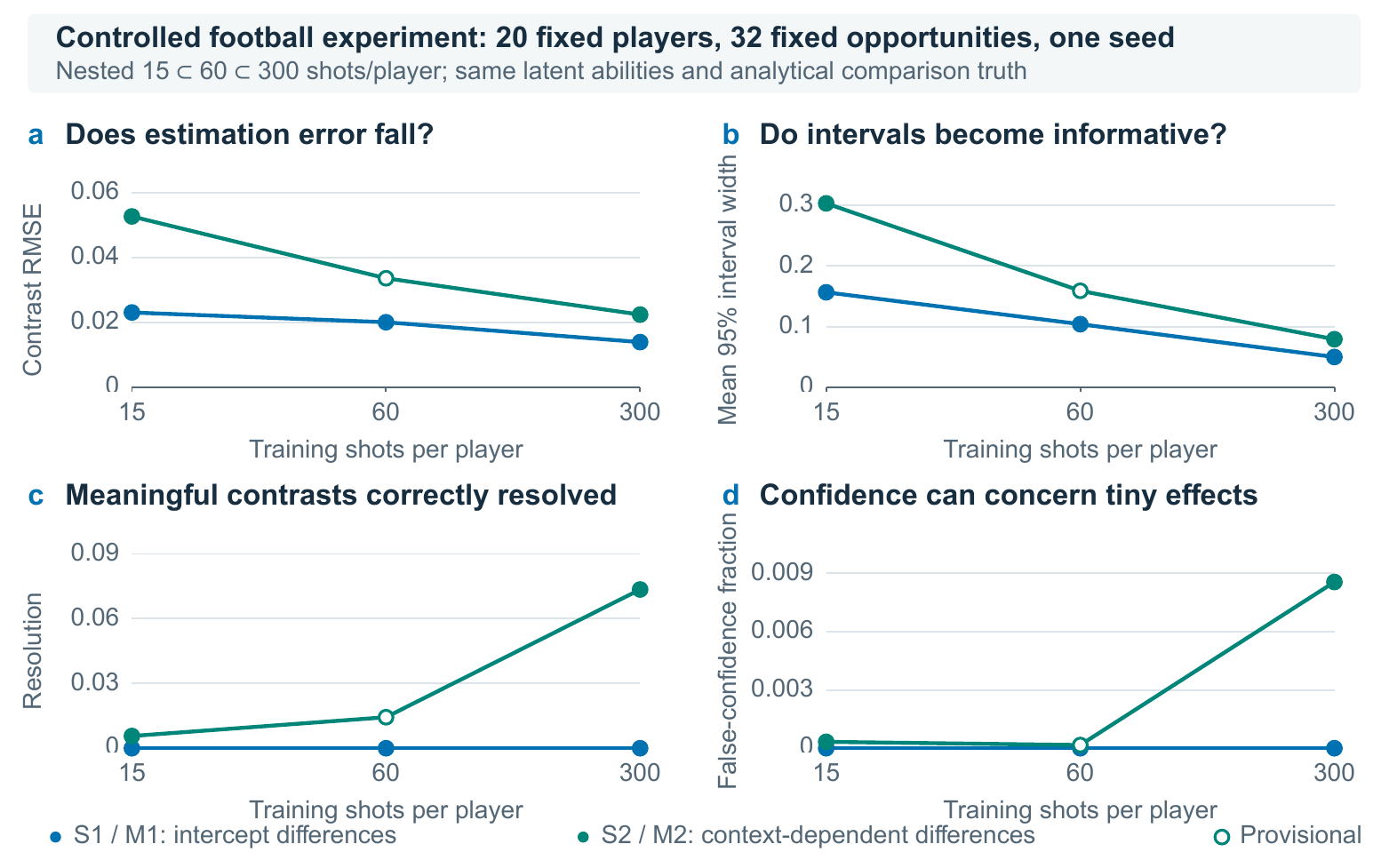}
\caption{
\textbf{More observations improve estimation without necessarily resolving player differences.}
The nested information experiment holds 20 simulated
players, their generating conversion effects and 32
evaluation opportunities fixed while increasing the
number of training shots from 15 to 60 and 300 per
player. Results cover 190 canonical seen--seen pairs
and 6,080 dependent pair/context cells at
$\epsilon=0.01$.
(a) Posterior-mean RMSE decreases with additional
observations.
(b) Mean equal-tail 95\% credible intervals become
narrower.
(c) Correct resolution of meaningful contrasts
increases under S2/M2 but remains zero under S1/M1.
(d) False directional confidence includes
wrong-direction claims and confident claims about
effects within $\pm\epsilon$. The hollow S2/60
marker denotes a provisional fit.
Although increasing information improves accuracy
and precision in this realisation, it does not
eliminate uncertainty about meaningful differences
or prevent confident conclusions about near-null
effects. The results do not establish population
calibration, a general sample-size relationship
or validated thresholds for football applications.
}
\label{fig:recovery}
\end{figure}

\subsection{Resolution, small differences and shrinkage}

The information experiment highlights a distinction
between learning more about a player and obtaining
sufficient evidence to distinguish that player from
another. Because the generating conversion mechanism
and evaluation opportunities remain fixed, the
changes across training sizes arise from the
additional observations available for inference.
The nested design also preserves earlier outcomes
and contexts, although the larger samples naturally
provide broader exposure to the generating
distribution of shooting circumstances.

The complementary recovery measures help explain
why these improvements do not occur at the same rate.
RMSE describes how closely posterior-mean estimates
approach the known differences, while credible
interval width reflects the concentration of
posterior uncertainty. Directional resolution
asks a more demanding question: whether the
available evidence supports a distinction from
zero in the correct direction. In S1, both
estimation accuracy and directional accuracy
improve substantially, but the meaningful
contrasts remain unresolved. Under S2, some
differences become identifiable, although
the majority remain inconclusive even with
300 observations per player.

These results also illustrate an important
consequence of partial pooling. Hierarchical
estimation can stabilise individual effects
when observations are sparse, reducing the
influence of unusually successful or
unsuccessful shooting records. At the same
time, shrinkage towards the population
distribution can make modest differences
difficult to distinguish, particularly
when the available evidence provides
limited information about individual
deviations. Increasing the number of
observations reduces some of this
uncertainty, but does not guarantee
that every genuine difference will
become confidently identifiable.

The interpretation of confident results
requires equal care. An interval that
excludes zero establishes a directional
distinction under the fitted model,
but it does not necessarily establish
that the difference exceeds a magnitude
of practical interest. The 52
false-confidence cells in S2/300
illustrate this problem. Most have
the correct sign, but their true
effects fall within the declared
near-null region; only two are
also incorrectly signed. These
outcomes represent different forms
of overinterpretation rather than
52 independent instances of
incorrect player ranking.
Because multiple cells share
players and evaluation contexts,
their errors are statistically
dependent.

The six-scenario pilot places
these findings within a broader
set of inferential difficulties.
S0 examines whether a model
unnecessarily differentiates
identical shooters, while S1
and S2 consider increasingly
flexible forms of genuine
heterogeneity. S3 introduces
unequal individual records,
reflecting the imbalance in
shooting opportunities commonly
encountered in observational
football data. S4 examines
the consequences of assigning
players to separated regions
of the distance distribution,
thereby creating comparisons
that depend more heavily on
unfamiliar shooting geometry.
Finally, S5 introduces a
specified nonlinear distance
effect absent from the fitted
M2 model, allowing us to
examine the consequences
of contextual misspecification.

These scenarios should be
interpreted as controlled
stress tests rather than
literal representations
of individual footballers.
In particular, the separated
distance exposure in S4
does not describe the role
of any named player, and
the quadratic term in S5
is a deliberately specified
departure from the fitted
model rather than a physical
theory of long-range finishing.
They nevertheless demonstrate
that uncertainty in player
comparison can arise from
several sources, including
limited observations,
unequal contextual exposure
and differences between
the generating and fitted
model structures. Sampling
convergence, interval
precision and successful
recovery therefore require
separate assessment.

Finally, the scope of
these experiments must
remain clear. Both use
random seed 2026, and
the evaluated contrasts
share simulated players,
model coefficients and
opportunity contexts.
The reported cells are
therefore dependent
observations within
single controlled
realisations, not
independent repetitions
from which population
coverage or general
sample-size requirements
can be established.
The empirical centring
of the intercept prior
also changes across
nested training subsets,
while finite-sample
centring of the generating
player effects differs
slightly from the
independent inference
hierarchy. These features
qualify the interpretation
of the observed
improvements without
altering the central
lesson of the experiment:
more information can
improve estimation
substantially while
leaving many individual
comparisons unresolved.

\section{Football applications on observed opportunity sets}
\label{sec:football}

We now apply the framework to historical shooting opportunities
involving four Premier League and Serie A players. These
applications illustrate how estimated differences between
shooters depend on the opportunities used for comparison,
and how those estimates should be interpreted alongside
posterior uncertainty and the players' observed exposure.

Three opportunity sets are considered: 110 eligible attempts
by Jamie Vardy, 102 by Olivier Giroud and 75 by Domenico
Berardi. Each set is defined by its original source event
identifiers, and both the factual and alternative shooter
are evaluated on every recorded context within that set.
The comparisons therefore hold the measured circumstances
constant within each direction, rather than attempting
to reconstruct what would have happened had another player
participated in the match. The factual goal counts refer
only to these eligible attempts and should not be confused
with official season totals. Appendix~\ref{app:reproducibility}
documents the event-level membership of each set and
distinguishes traceable observations from historical
correspondence that cannot be independently verified.

Table~\ref{tab:football} summarises the comparisons under
the primary player-intercept model M1 and the provisional
context-dependent extension M2. In each case, the contrast
represents the alternative shooter's expected-goal total
minus that of the factual shooter, evaluated on the same
opportunities. Positive values therefore favour the
alternative player within the specified set. The posterior
intervals and sign probabilities are essential to the
interpretation: a small difference between posterior means
does not, by itself, establish that the players have
equivalent conversion probabilities.

\begin{table}[tb]
\caption{
\textbf{Conditional conversion comparisons on fixed historical opportunities.}
Each direction evaluates two shooters on the same recorded
attempts. $\mathrm{V}\to\mathrm{G}$ assigns Giroud to Vardy's
110 opportunities, $\mathrm{G}\to\mathrm{V}$ assigns Vardy
to Giroud's 102 opportunities, and $\mathrm{B}\to\mathrm{S}$
assigns Sansone to Berardi's 75 opportunities. Goals are
the factual shooter's observed outcomes on eligible
attempts, while xG values are posterior means.
$\Delta$ denotes alternative minus factual expected goals,
with equal-tail 95\% credible intervals calculated from
paired posterior draws. Positive contrasts favour the
alternative shooter. M1 is the primary hierarchical model;
M2 remains provisional. All intervals include zero,
so none of the comparisons establishes a preferred
shooter.
}
\label{tab:football}
\centering
\scriptsize
\setlength{\tabcolsep}{3pt}
\begin{tabular}{llrrrlr}
\toprule
Set & Model & Goals & Factual xG & Alternative xG &
$\Delta$ [95\% CrI] & $\Prb(\Delta>0)$ \\
\midrule
$\mathrm{V}\to\mathrm{G}$ & M1 & 19 & 19.652 & 19.603 &
$-0.049$ [$-4.017,+3.850$] & 0.484 \\
 & M2 & 19 & 19.405 & 19.439 &
$+0.034$ [$-4.079,+4.415$] & 0.506 \\
$\mathrm{G}\to\mathrm{V}$ & M1 & 15 & 13.413 & 13.452 &
$+0.038$ [$-2.964,+3.117$] & 0.516 \\
 & M2 & 15 & 13.255 & 13.163 &
$-0.092$ [$-4.117,+3.720$] & 0.481 \\
$\mathrm{B}\to\mathrm{S}$ & M1 & 3 & 4.156 & 4.190 &
$+0.034$ [$-1.027,+1.210$] & 0.515 \\
 & M2 & 3 & 4.137 & 4.206 &
$+0.068$ [$-0.932,+1.152$] & 0.549 \\
\bottomrule
\end{tabular}
\end{table}

\subsection{Giroud--Vardy: two opportunity sets, one paired question}
\label{sec:gv}

The comparison between Giroud and Vardy provides a useful
illustration of why the opportunities on which players are
evaluated matter. Both were established Premier League
forwards during the 2015/16 season, yet their recorded
attempts differed in location and shooting geometry.
Rather than comparing their observed goal totals directly,
we ask how their estimated conversion probabilities
would differ if each were evaluated on the other's
recorded opportunities.

Vardy's eligible set contains 110 attempts, comprising
85 training-period and 25 validation-period contexts.
Giroud's contains 102 attempts, with 79 and 23 in the
respective periods. Under M1, evaluating Giroud on
Vardy's opportunities produces a posterior-mean
difference of $-0.049$ expected goals. Reversing the
exercise and evaluating Vardy on Giroud's opportunities
gives $+0.038$. Both differences are negligible relative
to their posterior uncertainty: the corresponding
95\% credible intervals are $[-4.017,+3.850]$ and
$[-2.964,+3.117]$, with sign probabilities of
0.484 and 0.516. The data therefore provide little
basis for preferring either shooter on the specified
opportunities.

The provisional M2 extension produces similarly
inconclusive results, although its posterior-mean
contrasts have the opposite signs. Assigning Giroud
to Vardy's attempts yields $+0.034$ expected goals,
whereas assigning Vardy to Giroud's gives $-0.092$.
These changes are small relative to the corresponding
intervals, both of which include zero. The apparent
reversal should therefore not be interpreted as
evidence that one player is better suited to a
particular collection of chances. Rather, it
illustrates how allowing player effects to vary
with context can change point estimates without
providing sufficient evidence to establish a
reliable difference.

\begin{figure}[tb]
\centering
\includegraphics[width=\linewidth]{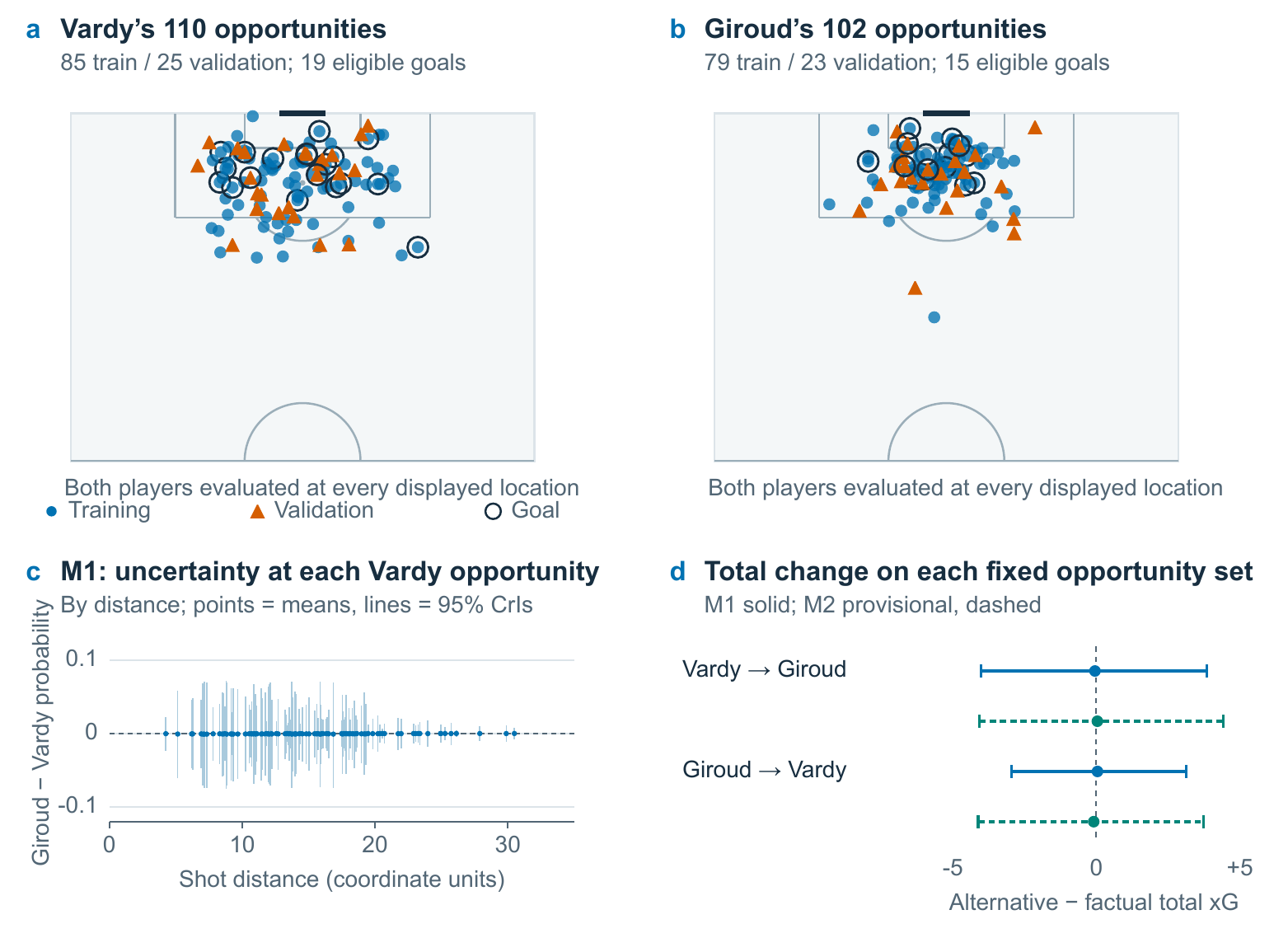}
\caption{
\textbf{Giroud and Vardy evaluated on each other's recorded opportunities.}
(a,b) Spatial distributions of Vardy's 110 and Giroud's
102 eligible attempts. Circles identify training-period
contexts, triangles validation-period contexts, and
rings observed goals. Both players are evaluated on
every opportunity within each set, although the two
sets contain different attempts. Half-pitches are
rotated counter-clockwise, placing the goal at the
top while preserving the source coordinates and
isotropic proportions.
(c) M1 paired probability differences across Vardy's
opportunities, plotted against shot distance. Points
show posterior means and lines indicate pointwise
equal-tail 95\% credible intervals.
(d) Aggregate expected-goal contrasts obtained by
summing differences within aligned posterior draws.
Solid blue intervals represent primary M1 and dashed
teal intervals provisional M2. Positive values favour
the alternative shooter on the specified opportunity
set. All aggregate intervals include zero. The
spatial panels describe observed shooting locations,
not a probability surface or a model of tactical
movement.
}
\label{fig:gv}
\end{figure}

The composition of the two opportunity sets helps
explain why comparisons should be evaluated in both
directions. Using distance quartiles derived from
the pooled training population, 59 of Giroud's
102 attempts fall within the closest quartile,
compared with 40 of Vardy's 110. A similar
difference appears in goal-opening angle:
65 of Giroud's attempts belong to the largest
opening-angle quartile, compared with 37 of
Vardy's. These distributions indicate that
the players encountered different collections
of shooting opportunities, although they do
not establish why those differences arose or
whether they reflect particular tactical roles.

Holding the opportunities fixed removes this
difference in observed composition from the
within-set comparison. Both shooters are
evaluated at the same locations and under
the same recorded circumstances, allowing
the estimated player contrast to be examined
without changing the contextual inputs.
Figure~\ref{fig:gv} illustrates the spatial
composition of the two sets and the uncertainty
associated with their paired comparisons.
The maps also distinguish training-period
opportunities, which contributed to model
estimation, from later validation-period
contexts.

The factual outcomes provide a useful
description of the historical records,
but they answer a different question.
Vardy scored 19 goals from his eligible
attempts, compared with an M1 factual
expectation of 19.652, while Giroud
scored 15 against 13.413. These
observed-minus-expected differences
reflect realised outcomes, opportunity
composition, player adjustment and
the inclusion of fitting-period
observations. They cannot be interpreted
as direct evidence of what either
player would have achieved on the
other's attempts.

The relationship between the two
directions also follows from the
model structure. Within any M1
posterior draw, the difference
between player intercepts determines
their probability ordering across
all contexts. Reversing the shooters
therefore reverses the sign of
that ordering, although the
aggregate magnitudes need not
be equal because the logistic
transformation makes probability
differences depend on opportunity
difficulty. M2 permits additional
variation through context-dependent
player effects, allowing the
composition of an opportunity
set to influence the ordering
within a draw.

Despite these modelling differences,
the substantive conclusion is
unchanged. Neither the historical
goal totals nor the posterior
comparisons provide sufficient
evidence to identify a preferred
shooter on these fixed sets of
opportunities. The estimated
differences are small, but the
credible intervals remain wide
enough to accommodate meaningful
advantages in either direction.

\subsection{Berardi--Sansone: geometry, pressure and unfamiliar opportunities}
\label{sec:bs}

The comparison between Berardi and Sansone extends
the analysis to a different distribution of
shooting opportunities and places greater
emphasis on contextual exposure. We evaluate
Sansone on Berardi's 75 eligible attempts,
of which 50 belong to the training period
and 25 to validation. Unlike the Giroud--Vardy
application, the comparison is presented
in one direction, using Berardi's recorded
opportunities as the common basis for
evaluation.

Under M1, Berardi's factual expected-goal
total is 4.156, compared with 4.190
for Sansone. The resulting contrast
of $+0.034$ expected goals has a
95\% credible interval of
$[-1.027,+1.210]$. M2 produces
a slightly larger posterior-mean
contrast of $+0.068$, with an
interval of $[-0.932,+1.152]$.
Although both posterior means
favour Sansone, neither interval
provides sufficient evidence
to distinguish the players'
conditional conversion
probabilities on these attempts.

The distribution of Berardi's
opportunities differs from
those considered in the
Premier League comparison.
Only five of his 75 attempts
fall within the closest
pooled-training distance
quartile, while 25 belong
to the furthest quartile.
His shooting angles are
also concentrated towards
narrower openings, with
26 and 25 attempts in
the two smallest
goal-opening quartiles.
These characteristics
describe the geometry
of his recorded attempts
without establishing
whether his shot selection
was more or less effective
than that of another player.
They also reinforce the
importance of interpreting
aggregate expected-goal
differences within the
particular opportunity
set on which they are
calculated.

Pressure annotations provide
a further opportunity to
examine how the comparison
behaves across recorded
circumstances. Of Berardi's
75 attempts, 21 carry a
pressure annotation.
On this subset, the M1
posterior-mean contrast
is $+0.011$ expected goals,
with a 95\% credible
interval of
$[-0.311,+0.369]$.
M2 produces a larger
mean difference of
$+0.059$, but its
interval of
$[-0.480,+0.760]$
remains inconclusive.

These estimates do not
establish that Sansone
would have performed
better under pressure.
The annotation identifies
recorded presence rather
than a controlled
experimental condition,
and the models do not
isolate pressure from
all other circumstances
that may influence
shooting outcomes.
Berardi scored no goals
on the 21 annotated-pressure
attempts and three on
the remaining 54, but
these small factual
counts cannot establish
that pressure prevented
goals or that an
alternative shooter
would have converted
the same chances.
The posterior comparison
addresses expected
conversion under fixed
recorded circumstances,
not the causal effect
of pressure or player
substitution.

\begin{figure}[tb]
\centering
\includegraphics[width=\linewidth]{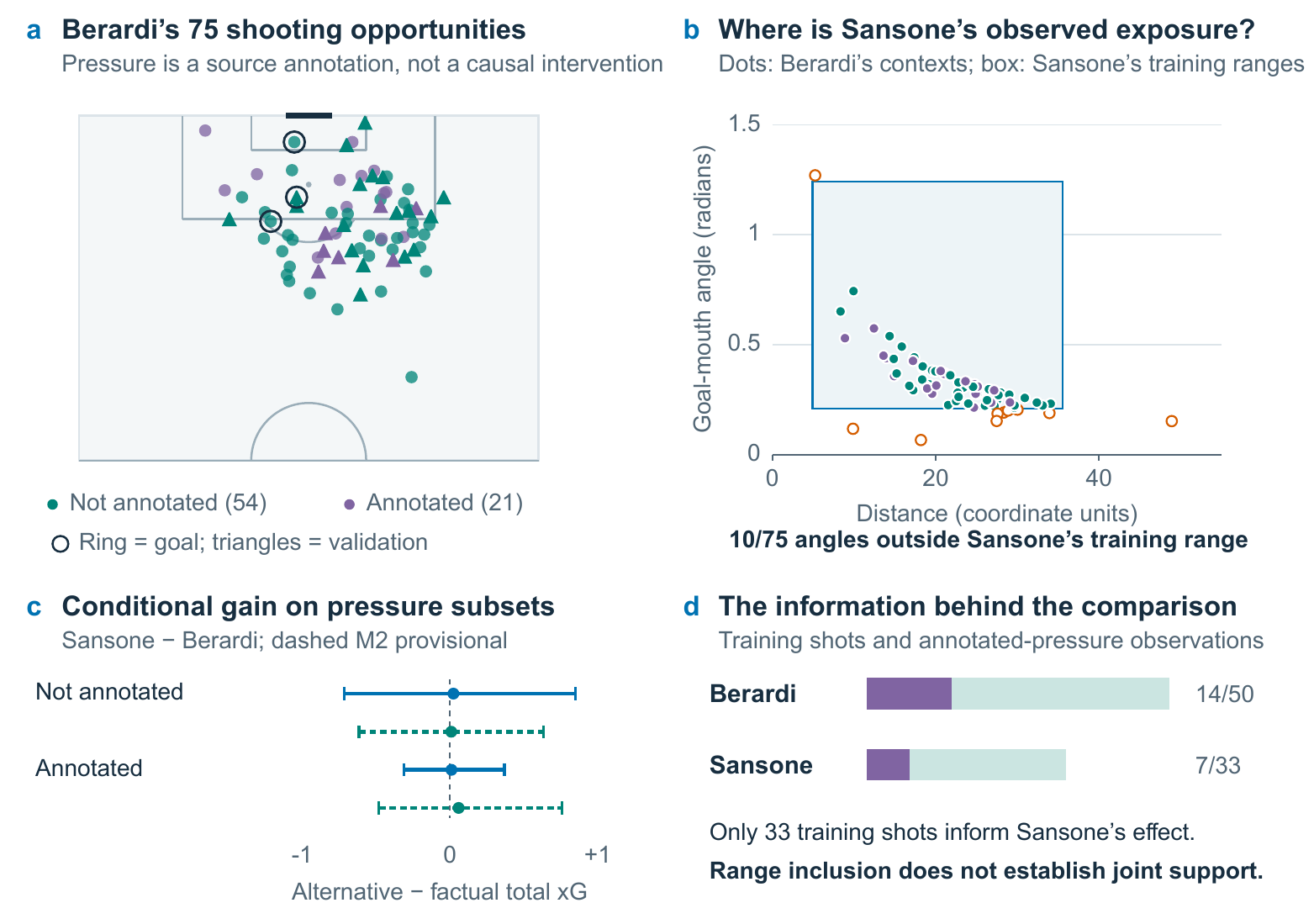}
\caption{
\textbf{Berardi--Sansone comparison across shooting geometry, pressure and observed exposure.}
(a) Berardi's 75 eligible opportunities displayed
on a counter-clockwise-rotated half-pitch,
with the goal at the top. Colour indicates
recorded pressure annotation, rings identify
factual goals, and triangles denote
validation-period contexts.
(b) The same attempts in distance--angle
space. The rectangle marks Sansone's
marginal training ranges, while hollow
orange edges identify opportunities
outside at least one range. Ten
attempts fall outside his angle
range and one outside his
distance range. The rectangle
does not establish joint
contextual support.
(c) Paired aggregate contrasts
for the 54 attempts without
recorded pressure and the
21 annotated-pressure attempts.
Solid blue intervals denote
primary M1 and dashed teal
intervals provisional M2.
Every equal-tail 95\%
credible interval includes
zero.
(d) Training shot counts
and recorded-pressure counts.
These panels describe
observed exposure and
conditional conversion
estimates; they do not
establish player
specialisation, differences
in composure or causal
effects of pressure.
}
\label{fig:bs}
\end{figure}

The interpretation of these
comparisons is further
complicated by differences
in the players' observed
training exposure. Sansone
contributes only 33 training
attempts, including seven
with recorded pressure.
Moreover, ten of Berardi's
opportunities fall outside
Sansone's observed training
angle range, while one
falls outside his distance
range (Fig.~\ref{fig:bs}).
These unfamiliar contexts
do not prevent the model
from generating predictions,
but they introduce an
additional reason for
caution when interpreting
the resulting contrasts.

We retain all 75 specified
opportunities rather than
excluding unfamiliar
contexts to obtain a
more favourable or
apparently precise
comparison. The exposure
summaries instead make
the extent of extrapolation
visible to the reader.
This distinction matters
because a fitted model
can assign a conversion
probability to almost
any valid contextual
input, even when the
corresponding player
has little or no
observed experience
of similar situations.

The Berardi--Sansone
application therefore
illustrates two related
limitations. First,
the available evidence
does not resolve the
difference between
the players on the
specified opportunities,
including those with
recorded pressure.
Second, the interpretation
of the posterior estimates
depends on contextual
exposure that cannot
be fully summarised
by the number of
recorded shots alone.

\subsection{Identifiability and contextual support across applications}
\label{sec:support}

The three comparisons reveal
a common difficulty:
uncertainty about player
differences must be
interpreted alongside
the circumstances in
which those players
have actually been
observed. A player
may contribute a
reasonable number
of attempts while
having little
experience of
particular combinations
of distance, angle,
pressure and
shooting technique.
Consequently, total
shot count provides
only a partial
description of
the information
available for
individual comparison.

The four application
players contribute
different amounts
of training evidence.
Vardy has 85
observations, Giroud
79, Berardi 50
and Sansone 33,
with pressure
annotated on 19,
29, 14 and seven
attempts, respectively.
Their observed
distance and angle
ranges are reported
in Table~\ref{tab:support}.
These summaries
describe the
circumstances
represented in
each player's
training record,
rather than
defining the
full range of
opportunities
that the player
could plausibly
convert.

\begin{table}[tb]
\caption{
\textbf{Observed training exposure for the four application players.}
Shot counts describe the eligible training attempts
available for estimating player effects, while
pressure counts indicate attempts with recorded
positive pressure annotations. Distance ranges
are expressed in StatsBomb coordinate units
and goal-opening angles in radians. The
reported minima and maxima describe marginal
exposure only; they do not establish whether
a player has sufficient observations near
a particular combination of contextual
features or provide validated measures
of joint support.
}
\label{tab:support}
\centering
\small
\setlength{\tabcolsep}{4pt}
\begin{tabular}{lrrll}
\toprule
Player & Shots & Pressure & Distance range & Angle range \\
\midrule
Vardy & 85 & 19 & 4.245--30.490 & 0.069--1.490 \\
Giroud & 79 & 29 & 4.494--35.263 & 0.152--1.448 \\
Berardi & 50 & 14 & 5.235--48.894 & 0.066--1.270 \\
Sansone & 33 & 7 & 4.920--35.574 & 0.209--1.241 \\
\bottomrule
\end{tabular}
\end{table}

The marginal ranges help
identify unfamiliar
opportunities in the
evaluated sets.
Giroud's training
distance range
excludes one of
Vardy's 110
contexts, while
his angle range
excludes four.
Vardy's distance
range excludes
two of Giroud's
102 contexts.
One of Giroud's
own validation
attempts also
falls outside
his training
angle range.

The differences
are more pronounced
for Sansone,
whose training
ranges exclude
one of Berardi's
75 attempts by
distance and ten
by angle. His
angle-range
inclusion is
therefore 86.7\%.
Both pressure
states occur
in the training
records of all
four players,
but this does
not imply that
each player
has substantial
experience of
every combination
of pressure
and shooting
geometry.

These observations
also show why
marginal ranges
should not be
treated as
definitive
measures of
contextual
support. An
opportunity
may fall within
a player's
observed
distance and
angle ranges
while occupying
a region of
their joint
distribution
that is scarcely
represented
in training.
Conversely,
a context
lying slightly
outside one
range may
still be
reasonably
similar to
observed
opportunities
in other
respects.

For this
reason, we
use the
exposure
summaries
descriptively
rather than
assigning
players or
opportunities
to arbitrary
categories
of reliability.
The available
data do not
support a
validated
threshold
for declaring
a comparison
well supported
or unsuitable
for interpretation.

Posterior
credible
intervals
and contextual
exposure
summaries
therefore
address
different
aspects of
uncertainty.
The intervals
describe
uncertainty
about model
parameters
and the
resulting
conversion
contrasts
under the
assumed
model and
priors.
Exposure
summaries
identify
regions
where
predictions
depend more
heavily on
information
shared across
players or
on extrapolation
beyond an
individual's
observed
record.

Neither
quantity
fully
substitutes
for the
other.
A narrow
posterior
interval
would not
necessarily
establish
that a
player's
contextual
experience
is sufficient,
while
observations
within
familiar
ranges
would not
guarantee
a precise
estimate.
M1 assumes
a common
relationship
between
shooting
geometry
and conversion,
with player
differences
represented
through
intercepts.
M2 allows
some of
those
relationships
to vary
between
players,
but doing
so requires
additional
individual
information
and does
not remove
the possibility
of extrapolation
or model
misspecification.

Finally,
the near-zero
posterior
means in
Table~\ref{tab:football}
should not
be interpreted
as evidence
that the
players are
equivalent.
The M1
credible
intervals
for the
larger
opportunity
sets
accommodate
differences
of several
expected
goals in
either
direction.
The evidence
is therefore
insufficient
to establish
superiority,
but it is
also
insufficient
to establish
equivalence.

A practical
decision
would require
a defensible
criterion
for the
magnitude
of a
meaningful
advantage,
together
with estimates
precise
enough to
evaluate
that
criterion.
The tolerance
values
used in
the synthetic
experiments
serve as
controlled
experimental
conventions
and should
not be
transferred
directly
to decisions
about these
players.

The applications
thus reinforce
the distinction
developed
throughout
the paper.
Evaluating
players
on identical
recorded
opportunities
makes their
estimated
conversion
differences
comparable,
but it
does not
guarantee
that the
available
evidence
can resolve
those
differences.
A responsible
interpretation
must consider
the posterior
contrast,
its
uncertainty
and the
contextual
information
supporting
the
prediction,
rather
than
reducing
the
comparison
to a
single
expected-goal
difference.

\section{Practical implications and limits of fixed-context comparison}
\label{sec:discussion}

The findings highlight an important distinction between predicting
the outcome of a shooting opportunity and identifying differences
between the players who might take it. The recorded circumstances
of an attempt provide substantial information about its probability
of resulting in a goal, as demonstrated by the improvement of all
three fitted models over the constant-rate benchmark. Yet the
additional contribution of player-specific effects is considerably
less certain. M1 offers only a marginal improvement in chronological
predictive performance over the contextual benchmark, while its
estimated differences between individual shooters remain small
relative to their posterior uncertainty. Although M2 allows those
differences to vary with shooting conditions, its sampling
qualification and inconclusive application intervals do not
support claims of established player specialisation. Reliable
prediction of observed outcomes can therefore coexist with
insufficient evidence to distinguish the players being compared.

The controlled experiments help explain why these findings are
not contradictory. Increasing the number of observations improves
estimation accuracy and, in some settings, directional accuracy,
but does not necessarily provide enough information to resolve
meaningful differences. Even as credible intervals become narrower,
many genuine contrasts remain uncertain, while some confidently
estimated differences concern effects smaller than the declared
tolerance. These outcomes demonstrate that estimation error,
interval coverage, precision and directional resolution describe
related but distinct properties of inference. A ranking based
solely on posterior means would conceal much of this information,
particularly when the estimated differences are small or their
uncertainty remains substantial.

The contribution of the framework is consequently not a new
ranking of footballers, but a more explicit account of what
can be learned from comparing their estimated conversion
probabilities. By evaluating players on identical recorded
opportunities and preserving the dependence between their
posterior predictions, the analysis separates differences
in observed opportunity composition from conditional
differences attributed to the shooter. It also makes clear
where that separation reaches its limits: a comparison may
be mathematically well defined without being sufficiently
identified by the available evidence.

\subsection{What an analyst can reasonably take from the results}

For an analyst, the starting point should be the opportunity
set itself. A comparison between two players has little
meaning unless the opportunities on which they are evaluated
are clearly specified and the reason for selecting them is
understood. Observed goals describe what happened in the
historical record, whereas factual and alternative-player
expected goals describe model-based expectations under the
recorded circumstances. The relevant comparison is therefore
the paired difference between those expectations, considered
alongside its posterior uncertainty and the players' observed
contextual exposure.

This distinction is particularly important when the
posterior means are close. In the Giroud--Vardy and
Berardi--Sansone applications, the estimated differences
are small, but their credible intervals remain sufficiently
wide to accommodate advantages in either direction.
The appropriate conclusion is not that the players
are equivalent, nor that one should be preferred on
the basis of a marginal numerical difference. Rather,
the available evidence does not resolve the comparison.
A practical preference would require both a defensible
criterion for what constitutes a meaningful advantage
and estimates precise enough to assess that criterion.

An inconclusive result can nevertheless be informative.
Opportunity maps establish which recorded situations
are being compared, while paired posterior evaluation
makes uncertainty in the difference explicit. Aggregating
within posterior draws preserves the dependence shared
across opportunities, avoiding an artificially simplified
assessment of the total. The controlled experiments further
show that additional observations may improve estimates
without producing a confident preference, and that
confidence in the sign of a difference does not necessarily
imply that its magnitude is important. Together, these
features provide a more disciplined basis for interpreting
player comparisons than a single expected-goal difference
or an ordering of posterior means.

The interpretation becomes more restricted when the
question shifts from conditional conversion to the
consequences of an actual substitution. Replacing a
player may change where and when shooting opportunities
arise, whether the incoming player receives the ball,
how teammates respond and how opponents defend.
Movement, availability, chance creation and shot
selection may all differ, but these processes are
not represented in the conversion likelihood.
Even a precisely estimated aggregate contrast would
therefore describe expected conversion on the specified
opportunities, not the change in team goals that
would result from replacing one footballer with another.
The framework is intended to clarify the former
question, not to resolve the latter.

\subsection{What remains unresolved}

Several limitations prevent the fitted player effects
from being interpreted as isolated measures of finishing
ability. Footballers are not randomly assigned to
shooting opportunities, and their observed records
may differ in ways that the available contextual
features do not capture. Shot placement, preceding
movement, preferred situations, team environment
and variation in event annotation may all contribute
to the estimated differences. Holding measured
circumstances fixed during prediction does not
establish that the players' historical estimation
records are otherwise comparable. The residual
player effects consequently remain conditional
statistical quantities rather than direct
measurements of an underlying finishing trait.

The relatively small fitted deviations admit
several explanations that cannot be distinguished
reliably with the present data. They may reflect
limited genuine heterogeneity, sparse goal outcomes,
partial pooling or aspects of the conversion
process that the model does not represent.
Hierarchical shrinkage can stabilise estimates
from small samples, but it can also attenuate
genuine differences. Although the information
experiment illustrates this trade-off, it does
not establish that every near-zero empirical
estimate is caused by shrinkage. The sampling
limitations of M2 introduce a separate concern,
particularly for the interpretation of
context-dependent player effects. Likewise,
the supplementary M0/M1 analyses demonstrate
consistency across the two examined prior
specifications without establishing general
prior robustness or showing that alternative
priors would resolve the identification problem.

The empirical evidence is also restricted
in its temporal and population coverage.
The analysis concerns one historical season
and a single chronological training--validation
partition, leaving the transferability of
the findings to other seasons and competitions
untested. The small number of goals recorded
by previously unseen players limits the
interpretation of subgroup predictive
performance. Source-presence annotations
and the requirement for complete contextual
features further restrict the measurement
and population coverage of the study.

Contextual support remains another unresolved
issue. The reported distance and angle ranges
provide a transparent description of marginal
exposure, but they cannot establish whether
an individual has sufficient experience of
particular combinations of shooting conditions.
A model may produce a precise-looking prediction
in a region where the player's observed
experience is limited, and posterior uncertainty
need not fully reflect the consequences of
contextual extrapolation or misspecification.
The single-seed recovery experiments and
provisional slope diagnostics similarly
limit conclusions about repeated-sampling
calibration and the identification of
context-dependent player differences.
Reproduction of the analysis is subject
to the data-version and archive
qualifications documented in
Appendix~\ref{app:reproducibility}.

These limitations suggest several priorities
for further investigation. Replicated nested
experiments would help establish how consistently
the observed recovery patterns arise across
different generated samples. More detailed
assessments of joint contextual support could
clarify when alternative-player predictions
depend heavily on unfamiliar opportunities.
Additional diagnostics and sensitivity analyses
would also be needed before the contextual
extension could support stronger interpretations
of individual specialisation.

Such developments should be directed towards
understanding when player differences can be
identified, rather than towards obtaining
a preferred ranking through changes to the
analysis. The present framework provides
a basis for this work by keeping three
questions distinct: how well a model predicts
observed shooting outcomes, what it estimates
when two players are evaluated on the same
opportunities, and what would be required
to infer the consequences of an actual
player substitution.

\section{Conclusion}
\label{sec:conclusion}

Comparing footballers on identical shooting opportunities
requires more than estimating the probability of a goal.
It requires an explicit treatment of uncertainty in the
difference between their conversion probabilities.
The hierarchical Bayesian framework developed here
provides such a comparison through paired posterior
evaluation, with within-draw aggregation extending
the analysis from individual attempts to total
expected goals.

Chronological validation demonstrates the predictive
value of recorded shooting circumstances, while
revealing only limited additional benefit from
player-specific effects. Controlled experiments
show that increasing information can improve
estimation without necessarily resolving meaningful
differences between shooters. The historical
Giroud--Vardy and Berardi--Sansone applications
reflect this difficulty: neither the primary
hierarchical model nor the provisional contextual
extension provides sufficient evidence to
establish a preferred player on the specified
opportunities.

The resulting comparisons are therefore most
useful when interpreted as conditional estimates,
accompanied by their uncertainty and the contextual
evidence supporting them. They do not establish
what would have happened had another player
taken the shot, nor do they predict the
consequences of tactical substitution or
transfer decisions. Their value lies in making
the distinction between prediction and
identification explicit, and in showing
why a well-defined comparison need not
yield a confident conclusion.

\appendix

\section{Reproducibility and supplementary result conventions}
\label{app:reproducibility}

The analyses are defined by the specified data versions,
chronological partitions, preprocessing rules and
synthetic generating mechanisms. Table~\ref{tab:data}
documents the mutually exclusive exclusions applied
to the source shot population, establishing the
eligible observations used throughout the empirical
analysis.

The real-data posterior arrays retain player effects
and source-event identifier mappings. For each model,
the 1,000 retained draws from four chains are flattened
in chain-major order, and factual and alternative-player
predictions are evaluated using identical posterior
indices. This alignment preserves the dependence
between predictions that share contextual parameters
and player effects. Computational checks confirm
zero self-contrasts, antisymmetry when shooters
are exchanged on the same contexts, finite
predicted probabilities and agreement between
summed opportunity-level differences and the
corresponding alternative-minus-factual
aggregate totals.

Distance and goal-opening angle summaries are
constructed using quartiles of the pooled
training covariate distribution. The distance
cutpoints are 12.243, 18.672 and 25.520
StatsBomb coordinate units, while the
corresponding angle cutpoints are 0.26191,
0.34042 and 0.53635 radians. Pressure
summaries retain the source-presence encoding
described in Section~\ref{sec:data}.
Subgroup intervals are correlated and
pointwise, without adjustment for multiple
comparisons. They are therefore presented
as descriptive summaries rather than
independently validated evidence of
player specialisation.

The empirical opportunity sets comprise
287 unique source event identifiers across
the three applications. Per-shot conversion
probabilities, paired differences, sign
probabilities, contextual groupings and
support summaries are indexed by these
identifiers, ensuring that the reported
quantities refer to the same recorded
attempts. Under M1, the posterior
probabilities of a negative aggregate
contrast are 0.516 for Vardy-to-Giroud,
0.484 for Giroud-to-Vardy and 0.485
for Berardi-to-Sansone. No posterior
draw produces an exactly zero
aggregate difference. Training-period
and validation-period summaries use
the same posterior fitted to the
training data, rather than separately
estimated models.

The synthetic analysis comprises the
initial 18-fit scenario matrix and
six nested information conditions.
Sampling qualifications are retained
alongside the corresponding recovery
results. The primary tolerance for
meaningful per-shot probability
differences is 0.01, with sensitivity
analyses at 0.005 and 0.02.
For aggregate comparisons, the
tolerance is multiplied by the
32 evaluation opportunities.

In the S2 experiment with 300
training shots per player, two
incorrectly positive 95\%
contrast intervals concern the
same player pair evaluated in
two different contexts. Both
true differences have absolute
magnitudes below 0.01.
These cases illustrate how
confident directional claims
can arise for near-null
effects, but they are
dependent observations
rather than independent
replications from which
uncertainty in a recovery
rate could be estimated.

\subsection{Figure construction and evidential boundaries}

The vector figures combine recorded
shot locations and freeze-frame
information with predictive
evaluations, synthetic recovery
results and paired posterior
comparisons. Blue and teal
identify model comparisons,
while orange denotes
validation-period or
outside-range contexts
where indicated. Purple
identifies pressure or
goalkeeper positions
in the relevant panels.

Shot locations are displayed
without jittering or
smoothing. Rings identify
factual goals, while
period-specific symbols
distinguish the contexts
used for training from
those reserved for
validation. These
symbols do not indicate
different fitted models.
The hypothetical
assignments in
Fig.~\ref{fig:concept}
illustrate the
evaluation of
alternative shooters
without generating
counterfactual
goal outcomes.

Probability panels,
aggregate contrasts
and subgroup intervals
are calculated from
the paired posterior
draws underlying
the numerical results.
Chain-major indexing
maintains alignment
between the two
shooters' evaluations,
and distance and
angle calculations
were independently
checked against
source shot locations.
The freeze frame
in Fig.~\ref{fig:difficulty}
corresponds to the
first source event
in Vardy's opportunity
set. Figure~\ref{fig:concept}
uses the retrospective,
range-restricted
selection specified
in its caption.
Consequently, the
figures and tables
describe the same
quantities for the
same recorded
opportunities.

Source event identifiers
define membership of
the empirical opportunity
sets. The event-ID list
records each included
shot, its factual
shooter and match,
and its membership
of the training
or validation period.
These identifiers
link the analysis
to eligible StatsBomb
events at the
specified repository
revision.

Together with the
locally available
partition, preprocessing
configuration and
fitted posterior
arrays, this information
allows the present
comparisons to be
traced and reconstructed.
The historical
shot lists used in
the earlier
Giroud--Vardy and
Berardi--Sansone
case studies are
not available,
however, and their
individual events
cannot be matched
against those
used here.
Agreement in
the number of
shots therefore
does not establish
event-for-event
correspondence
between the
studies.

Finally, although
the underlying
event data remain
publicly available,
no public archive
containing the
complete study
code, fitted
posterior arrays
and derived
outputs has
been established.
The available
materials therefore
provide local
traceability
rather than a
fully reproducible
public research
package.

\begingroup\renewcommand{\bibfont}{\small}\setlength{\bibsep}{2pt}
\interlinepenalty=10000
\bibliographystyle{SageH}
\bibliography{references}

@article{code_2,
  author       = {Scholtes, A. and Karaku{\c{s}}, O.},
  title        = {{Bayes-xG}: player and position correction on expected goals ({xG}) using {Bayesian} hierarchical approach},
  journal      = {Frontiers in Sports and Active Living},
  year         = {2024},
  volume       = {6},
  pages        = {1348983},
  doi          = {10.3389/fspor.2024.1348983}
}

@article{code_3,
  author       = {Hewitt, J. H. and Karaku{\c{s}}, O.},
  title        = {A machine learning approach for player and position adjusted expected goals in football (soccer)},
  journal      = {Franklin Open},
  year         = {2023},
  volume       = {4},
  pages        = {100034},
  doi          = {10.1016/j.fraope.2023.100034}
}

@article{code_8,
  author       = {Mead, J. and O'Hare, A. and McMenemy, P.},
  title        = {Expected goals in football: Improving model performance and demonstrating value},
  journal      = {PLOS One},
  year         = {2023},
  volume       = {18},
  pages        = {e0282295},
  doi          = {10.1371/journal.pone.0282295}
}

@article{code_15,
  author       = {Rubin, D. B.},
  title        = {Estimating causal effects of treatments in randomized and nonrandomized studies},
  journal      = {Journal of Educational Psychology},
  year         = {1974},
  volume       = {66},
  pages        = {688--701},
  doi          = {10.1037/h0037350}
}

@book{code_16,
  author       = {Pearl, J.},
  title        = {Causality: Models, Reasoning and Inference},
  publisher    = {Cambridge University Press},
  address      = {Cambridge},
  year         = {2000}
}

@misc{code_24,
  author       = {StatsBomb},
  title        = {Free football data from {StatsBomb}},
  howpublished = {GitHub},
  year         = {n.d.},
  note         = {Available at: \url{https://github.com/statsbomb/open-data} (accessed 13 October 2026)}
}

@article{anzer_bauer_2021,
  author       = {Anzer, Gabriel and Bauer, Pascal},
  title        = {A Goal Scoring Probability Model for Shots Based on Synchronized Positional and Event Data in Football (Soccer)},
  journal      = {Frontiers in Sports and Active Living},
  year         = {2021},
  volume       = {3},
  pages        = {624475},
  doi          = {10.3389/fspor.2021.624475}
}

@article{hoffman2014,
 author={Hoffman, Matthew D. and Gelman, Andrew},
 title={The {No-U-Turn} Sampler: Adaptively Setting Path Lengths in {Hamiltonian Monte Carlo}},
 journal={Journal of Machine Learning Research}, year={2014}, volume={15}, number={47}, pages={1593--1623},
 url={https://www.jmlr.org/papers/v15/hoffman14a.html}
}

@article{pymc2023,
 author={Abril-Pla, Oriol and others},
 title={{PyMC}: a modern, and comprehensive probabilistic programming framework in {Python}},
 journal={PeerJ Computer Science}, year={2023}, volume={9}, pages={e1516}, doi={10.7717/peerj-cs.1516}
}

@article{arviz2019,
 author={Kumar, Ravin and Carroll, Colin and Hartikainen, Ari and Martin, Osvaldo},
 title={{ArviZ} a unified library for exploratory analysis of {Bayesian} models in {Python}},
 journal={Journal of Open Source Software}, year={2019}, volume={4}, number={33}, pages={1143}, doi={10.21105/joss.01143}
}

@article{vehtari2021,
 author={Vehtari, Aki and Gelman, Andrew and Simpson, Daniel and Carpenter, Bob and B{\"u}rkner, Paul-Christian},
 title={Rank-normalization, folding, and localization: An improved {R-hat} for assessing convergence of {MCMC}},
 journal={Bayesian Analysis}, year={2021}, volume={16}, number={2}, pages={667--718}, doi={10.1214/20-BA1221}
}

@article{code_1,
  author = {Herold, Mat and Goes, Floris and Nopp, Stephan and Bauer, Pascal and Thompson, Chris and Meyer, Tim},
  title = {Machine learning in men’s professional football: Current applications and future directions for improving attacking play},
  year = {2019},
  doi = {10.1177/1747954119879350},
  journal = {International Journal of Sports Science \& Coaching},
  volume = {14},
  number = {6},
  pages = {798--817}
}

@inproceedings{code_9,
  author = {Cavus, Mustafa and Biecek, Przemys\l{}aw},
  title = {Explainable expected goal models for performance analysis in football analytics},
  year = {2022},
  doi = {10.1109/DSAA54385.2022.10032440},
  booktitle = {2022 IEEE 9th International Conference on Data Science and Advanced Analytics (DSAA)},
  pages = {1--9},
  address = {Shenzhen, China, 13--16 October 2022}
}

@article{code_10,
  author = {Bandara, Ishara and Shelyag, Sergiy and Rajasegarar, Sutharshan and Dwyer, Dan and Kim, Eun-jin and Angelova, Maia},
  title = {Predicting goal probabilities with improved {xG} models using event sequences in association football},
  year = {2024},
  doi = {10.1371/journal.pone.0312278},
  journal = {PLOS ONE},
  volume = {19},
  number = {10},
  pages = {e0312278}
}

@article{code_11,
  author = {Santos-Fernandez, Edgar and Wu, Paul and Mengersen, Kerrie L.},
  title = {Bayesian statistics meets sports: a comprehensive review},
  year = {2019},
  doi = {10.1515/jqas-2018-0106},
  journal = {Journal of Quantitative Analysis in Sports},
  volume = {15},
  number = {4},
  pages = {289--312}
}

@article{code_12,
  author = {Baio, Gianluca and Blangiardo, Marta},
  title = {Bayesian hierarchical model for the prediction of football results},
  year = {2010},
  doi = {10.1080/02664760802684177},
  journal = {Journal of Applied Statistics},
  volume = {37},
  number = {2},
  pages = {253--264}
}

@article{code_19,
  author = {Nakahara, Hiroshi and Takeda, Kazuya and Fujii, Keisuke},
  title = {Estimating the effect of hitting strategies in baseball using counterfactual virtual simulation with deep learning},
  year = {2023},
  doi = {10.2478/ijcss-2023-0001},
  journal = {International Journal of Computer Science in Sport},
  volume = {22},
  number = {1},
  pages = {1--12}
}

@article{code_22,
  author = {Wang, Zhe and Veli\v{c}kovi\'{c}, Petar and Hennes, Daniel and Toma\v{s}ev, Nenad and Prince, Laurel and Kaisers, Michael and Bachrach, Yoram and Elie, Romuald and Wenliang, Li Kevin and Piccinini, Federico and Spearman, William and Graham, Ian and Connor, Jerome and Yang, Yi and Recasens, Adri\`{a} and Khan, Mina and Beauguerlange, Nathalie and Sprechmann, Pablo and Moreno, Pol and Heess, Nicolas and Bowling, Michael and Hassabis, Demis and Tuyls, Karl},
  title = {{TacticAI}: An {AI} assistant for football tactics},
  year = {2024},
  doi = {10.1038/s41467-024-45965-x},
  journal = {Nature Communications},
  volume = {15},
  number = {1},
  pages = {1906}
}

@article{brechot2020,
  author = {Brechot, Marc and Flepp, Raphael},
  title = {Dealing With Randomness in Match Outcomes: How to Rethink Performance Evaluation in {European} Club Football Using Expected Goals},
  year = {2020},
  doi = {10.1177/1527002519897962},
  journal = {Journal of Sports Economics},
  volume = {21},
  number = {4},
  pages = {335--362}
}

@article{waghmare2026,
  author = {Waghmare, Kartik and Ziegel, Johanna},
  title = {Proper Scoring Rules for Estimation and Forecast Evaluation},
  year = {2026},
  doi = {10.1146/annurev-statistics-042424-050626},
  journal = {Annual Review of Statistics and Its Application},
  volume = {13},
  number = {1},
  pages = {271--296}
}

@article{vancalster2019,
  author = {Van Calster, Ben and McLernon, David J. and van Smeden, Maarten and Wynants, Laure and Steyerberg, Ewout W.},
  title = {Calibration: the {Achilles} heel of predictive analytics},
  year = {2019},
  doi = {10.1186/s12916-019-1466-7},
  journal = {BMC Medicine},
  volume = {17},
  number = {1},
  pages = {230},
  note = {On behalf of the STRATOS initiative Topic Group on evaluating diagnostic tests and prediction models}
}

@article{gabry2019,
  author = {Gabry, Jonah and Simpson, Daniel and Vehtari, Aki and Betancourt, Michael and Gelman, Andrew},
  title = {Visualization in {Bayesian} workflow},
  year = {2019},
  doi = {10.1111/rssa.12378},
  journal = {Journal of the Royal Statistical Society Series A: Statistics in Society},
  volume = {182},
  number = {2},
  pages = {389--402}
}

@misc{code_5,
  author = {{Hudl Statsbomb}},
  title = {{StatsBomb} data case studies: Freeze frames and defender locations},
  year = {2021},
  howpublished = {In: Hudl Statsbomb, 3 March},
  note = {Available at: \url{https://blogarchive.statsbomb.com/news/statsbomb-data-case-studies-freeze-frames-and-defender-locations/} (accessed 13 October 2026)}
}

@misc{code_6,
  author = {{Get Goalside}},
  title = {So... everyone has pressure data now},
  year = {2021},
  howpublished = {In: Get Goalside, 25 November},
  note = {Available at: \url{https://www.getgoalsideanalytics.com/so-everyone-has-pressure-data-now/} (accessed 13 October 2026)}
}

@inproceedings{code_7,
  author = {Lucey, Patrick and Bialkowski, Alina and Monfort, Mathew and Carr, Peter and Matthews, Iain},
  title = {{``Quality vs Quantity''}: Improved shot prediction in soccer using strategic features from spatiotemporal data},
  year = {2015},
  booktitle = {{MIT} Sloan Sports Analytics Conference},
  pages = {1--9},
  note = {Available at: \url{https://www.sloansportsconference.com/research-papers/quality-vs-quantity-improved-shot-prediction-in-soccer-using-strategic-features-from-spatiotemporal-data} (accessed 13 October 2026)}
}

@inproceedings{code_13,
  author = {Egidi, Leonardo and Gabry, Jonah S.},
  title = {Bayesian hierarchical models for predicting individual performance in football (soccer)},
  year = {2017},
  booktitle = {MathSport International 2017 Conference Proceedings},
  pages = {112--121},
  note = {Available at: \url{https://www.mathsportinternational.com/MathSport2017Proceedings.pdf} (accessed 13 October 2026)}
}

@misc{code_21,
  author = {Alam, Shomoita and Moodie, Erica E. M. and Wu, Lucas Y. and Swartz, Tim B.},
  title = {Framing causal questions in sports analytics: A tutorial on estimand choice illustrated through crossing in soccer},
  year = {2026},
  howpublished = {arXiv preprint arXiv:2505.11841v2, 24 July},
  doi = {10.48550/arXiv.2505.11841},
  note = {Available at: \url{https://arxiv.org/abs/2505.11841} (accessed 13 October 2026)}
}
\endgroup
\end{document}